\documentclass{jfm}
\usepackage{graphicx}
\usepackage{epstopdf, epsfig}
\usepackage{amsfonts,amssymb,amsmath,mathbbol}   

\usepackage{xcolor} 

\def\We{{\it We}}

\shorttitle{Drop size distribution in an airstream}
\shortauthor{S. S. Ade, L. D. Chandrala and K. C. Sahu}

\title{Size distribution of a drop undergoing breakup at moderate Weber numbers}

\author{Someshwar Sanjay Ade,\aff{1} Lakshmana Dora Chandrala\aff{2} \and \\ Kirti Chandra Sahu\aff{3}\corresp{\email{lchandrala@mae.iith.ac.in, ksahu@che.iith.ac.in}}}

\affiliation{\aff{1} Center for Interdisciplinary Program, Indian Institute of Technology Hyderabad, Kandi - 502 284, Sangareddy, Telangana, India
\aff{2}Department of Mechanical and Aerospace Engineering, Indian Institute of Technology Hyderabad, Kandi - 502 284, Sangareddy, Telangana, India
\aff{3}Department of Chemical Engineering, Indian Institute of Technology Hyderabad, Kandi - 502 284, Sangareddy, Telangana, India}

\begin{document}

\maketitle

\begin{abstract}
The size distribution of child droplets resulting from a dual-bag fragmentation of a water drop is investigated using shadowgraphy and digital in-line holography techniques. It is observed that parent drop fragmentation contributes to the atomization of tiny child droplets, while core drop disintegration predominantly results in larger fragments. Despite the complexity associated with dual-bag fragmentation, we demonstrate that it exhibits a bi-modal size distribution. In contrast, the single-bag breakup undergoes a tri-modal size distribution. We employ the analytical model developed by \cite{jackiw2022prediction} for dual-bag fragmentation that convincingly predicts the experimentally observed droplet volume probability density. We also estimate the temporal evolution of child droplet production in order to quantitatively illustrate the decomposition into initial and core breakups. Furthermore, we confirm that the analytical model adequately predicts the droplet size distribution for a range of Weber numbers.
\end{abstract}


\vspace{-1.5cm}

\section{Introduction} \label{sec:intro}
Droplet fragmentation and resulting child droplets size distribution (DSD) are important for a wide range of industrial applications, such as combustion, surface coating, pharmaceutical production, disease transmission modelling, and artificial rain technology, to name a few \citep{villermaux2007fragmentation,xu2022droplet,raut2021microphysical}. They are also essential for understanding natural phenomena like clouds and rainfall \citep{Villermaux2009single,villermaux2011distribution}. 

In an airstream, a drop undergoes different breakup modes, such as vibrational, bag, bag-stamen, multi-bag, shear, and catastrophic breakup modes, due to the competition between the aerodynamic and surface tension forces \citep{pilch1987use,guildenbecher2009secondary,suryaprakash2019secondary,soni2020deformation}. The Weber number, defined as $\We \equiv \rho_a U^2 d_0/\sigma$, is used to characterise the droplet breakup phenomenon in a continuous airstream. Here, $\rho_a$, $\sigma$, $U$, and $d_0$ denote the air density, interfacial tension, average velocity of the airstream, and equivalent spherical diameter of the drop, respectively. The critical Weber number $(\We_{cr})$ at which the transition from the vibrational to the bag breakup occurs is about 12 and 6 in the cross-flow \citep{taylor1963shape,kulkarni2014bag,soni2020deformation} and oppose-flow \citep{Villermaux2009single} configurations, respectively. For the intermediate Weber numbers ($28 \le \We \le 41$), \cite{cao2007new} were the first to observe dual-bag breakup mode by performing shadowgraph and high-speed imaging. \cite{boggavarapu2021secondary} investigated fragmentation of water and surrogate fuels droplets and observed bag, bag-stamen, dual-bag, and multi-bag breakup modes at different Weber numbers. Catastrophic fragmentation occurs when the drop explodes into a cluster of tiny fragments at very high Weber numbers. Recently, \cite{kirar2022experimental} studied the fragmentation of a drop under a swirl airstream using the shadowgraphy technique for a fixed Weber number. They found a new breakup mechanism termed as ``retracting bag breakup" for intermediate swirl strength. In this breakup mode, the ligaments are stretched in opposite directions by the swirling airstream, causing the drop to undergo capillary instability and breakup. The multi-bag breakup mode is the most challenging of all breakup modes since the development and breakup of bags happen at different times. Moreover, different parts, such as bags, nodes, and rim, of the drop undergo various breakup mechanisms simultaneously.

Few researchers \citep{gao2013quantitative,guildenbecher2016high,guildenbecher2017characterization,essaidi2021aerodynamic,jackiw2022prediction,ade2022droplet,li2022secondary} have investigated the DSD generated by different breakup mechanisms. By developing an analytical model for the combined multi-modal distribution, \cite{jackiw2022prediction} predicted the experimental results of \cite{guildenbecher2017characterization} for single-bag and sheet-thinning breakups. Digital in-line holography has recently emerged as a powerful tool to estimate the droplet size distribution \citep{shao2020machine,radhakrishna2021experimental,ade2022droplet,gao2013quantitative,guildenbecher2016high,guildenbecher2017characterization,essaidi2021aerodynamic}. \cite{radhakrishna2021experimental} investigated the effect of the Weber number on droplet fragmentation at high Ohnesorge numbers. They examined different breakup modes and reported DSD obtained using  the digital in-line holography technique. Using the digital in-line holography technique, \cite{ade2022droplet} showed that while the fragmentation results in mono-modal size distribution for the no-swirl airstream, it exhibits bi-modal and multi-modal distributions in a swirl flow for the low and high swirl strengths, respectively. They also showed the temporal variation of the DSD during fragmentation. For the swirling airstream, they implemented a theoretical analysis accounting for various mechanisms, such as the nodes, rim, and bag breakup mode, that predicted the experimentally obtained DSD for different swirl strengths. \cite{boggavarapu2021secondary} investigated the DSD for different breakup modes using the particle/droplet image analysis (PDIA) technique. They found that while the bag and bag-stamen breakups undergo tri-modal size distribution, the dual-bag and multi-bag breakups produce bi-model droplet size distribution. However, this study provides the DSD only at the final instant of the fragmentation process.

In the present study, we investigate the dual-bag fragmentation of a water drop and the temporal evolution of the DSD using shadowgraphy and digital in-line holography techniques. We employ a deep-learning-based post-processing method to capture three-dimensional information about an object with a high spatial resolution and also offers the spatial distribution of child droplets. An analytical model is also presented that satisfactorily predicts the experimentally observed droplet volume probability density. To the best of our knowledge, the present study is the first to utilize digital in-line holography and convolutional neural networks (CNN) to estimate the DSD of a water drop undergoing dual-bag fragmentation. Moreover, we employed a deep-learning-based image processing technique for the segmentation of child droplets. We use the analytical model developed by \cite{jackiw2022prediction} to predict the droplet volume probability density associated with single-bag, dual-bag, and multi-bag fragmentations. We found that our experimental results agree with the prediction of the analytical model that shows a bi-model distribution for the dual-bag breakup.

\section{Experimental set-up}
\label{sec:expt}
\begin{figure}
\centering
\includegraphics[width=0.8\textwidth]{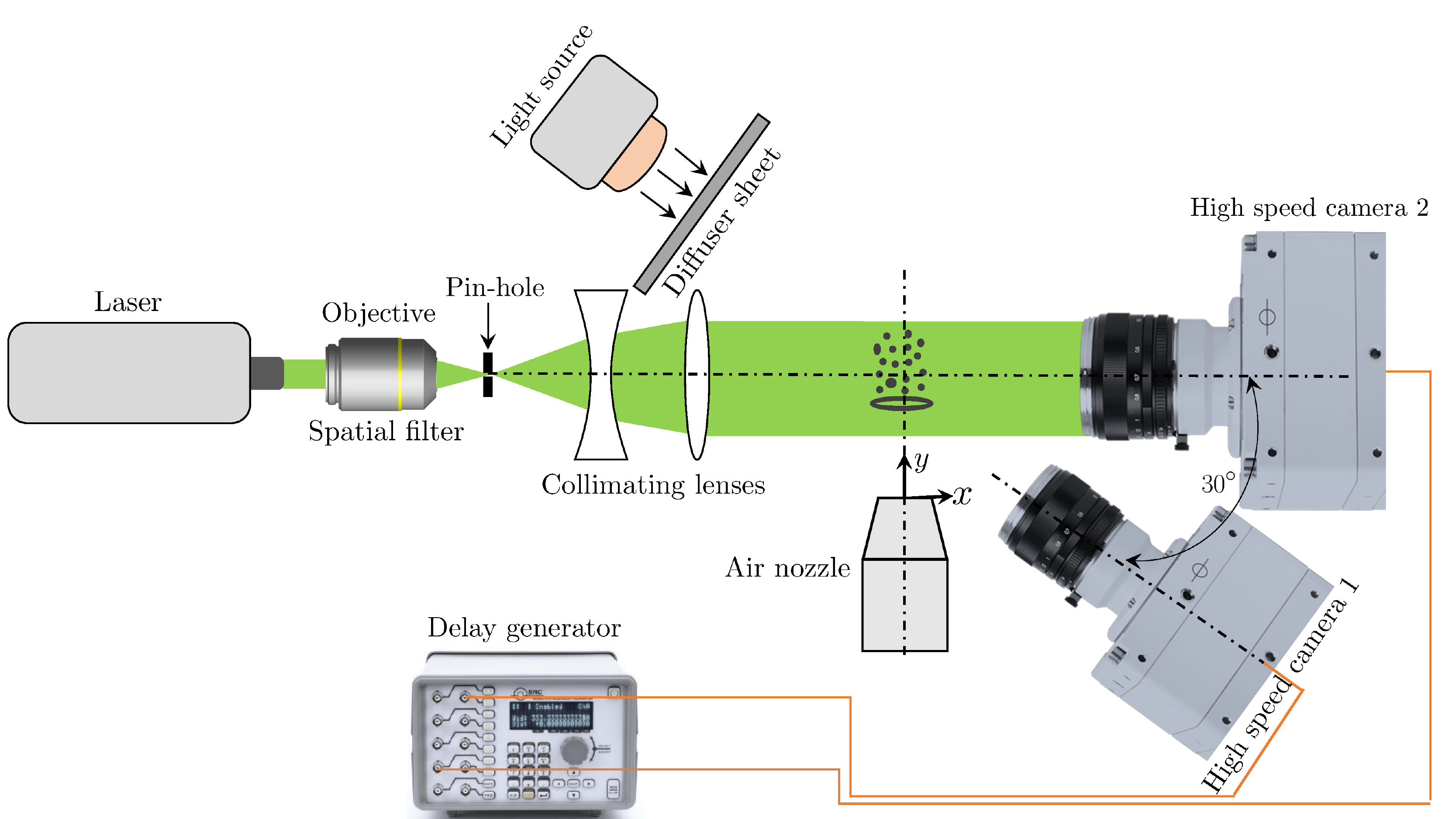}
\caption{Schematic diagram of the experimental set-up (top view).}
\label{fig1}
\end{figure}

A schematic diagram of the experimental set-up is shown in figure \ref{fig1}. It consists of (i) an air nozzle (with an inner diameter, $D_n=18$ mm) connected with a digital mass flow controller (model: MCR-500SLPM-D/CM, Make: Alicat Scientific, Inc., USA) and an air compressor, (ii) a droplet dispensing needle (20 Gauge), (iii) a continuous wave laser (model: SDL-532-100T, make: Shanghai Dream Lasers Technology Co. Ltd.), (iv) a spatial filter arrangement and collimating optics (make: Holmarc Opto-Mechatronics Ltd.), (v) two high-speed cameras (model: Phantom VEO 640L; make: Vision Research, USA) synchronized using a digital delay generator (model: 610036, Make: TSI, USA) and (vi) a diffused backlit illumination. 

A Cartesian coordinate system $(x,y,z)$ with its origin at the center of the nozzle is used to describe the dynamics. The dispensing needle is located at $(x_{d}/D_{n}, y_{d}/D_{n}, z_{d}/D_{n}) = (0, 0.35, 1.08)$. Except for $\We=40.15$ (discussed in \S\ref{highWe}), in the rest of our study, a water drop of diameter $d_{0}=3.09 \pm 0.07$ mm is injected from the needle. For $\We=40.15$, $d_0=4.3 \pm 0.07$ mm. The dimensionless time is defined as $\tau=Ut\sqrt{\rho _a/\rho_w}/d_{0}$, such that $\tau = 0$ represents the onset of breakup. Here, $t$ is dimensional time in second and $\rho_w = 998$ kg m$^{-3}$ is the density of water. The high-speed camera 1 with a Nikkor lens (focal length of 135 mm and minimum aperture of $f/2$) is employed for shadowgraphy. This camera is positioned at $x=180$ mm with an angle of $-30^\circ$ to the $x$ axis. A high-power light-emitting diode (model: MultiLED QT, Make: GSVITEC, Germany) is used along with a uniform diffuser sheet to illuminate the background. The resolution of the images captured using the high-speed camera 1 at 1800 frames per second (fps) with an exposure duration of 1 $\mu$s and a spatial resolution of 31.88 $\mu$m/pixel is $2048 \times 1600$ pixels. 

\begin{figure}
\centering
\includegraphics[width=0.8\textwidth]{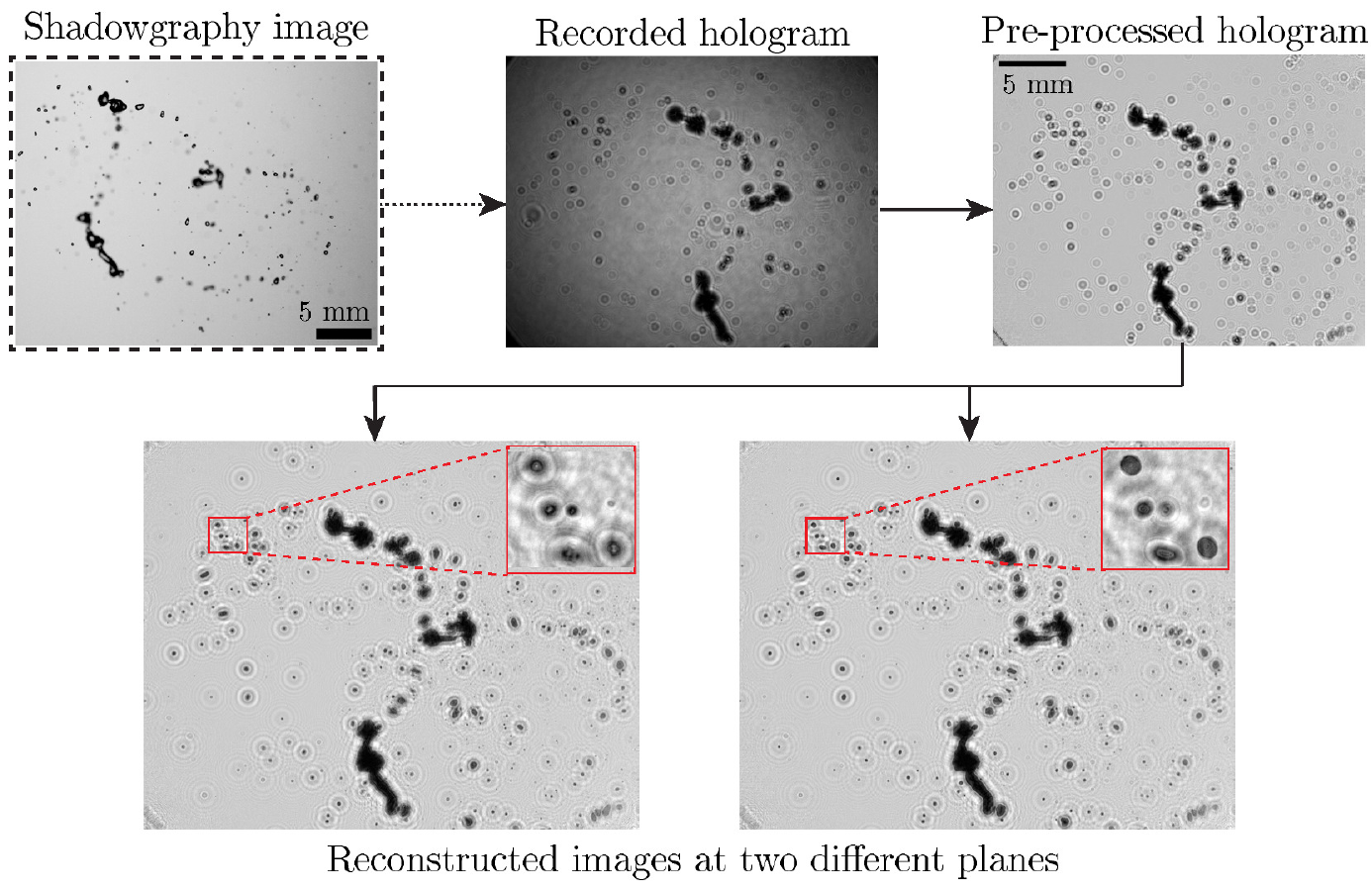}
\caption{Demonstration of different steps of the in-line holography. The first panel shows a typical shadowgraphy image of the satellite droplets during the fragmentation process. The second and third panels show recorded and pre-processed holograms. The bottom panels depict the reconstructed images at two different planes.}
\label{figS2}
\end{figure}

A continuous wave laser (output power 100 mW and wavelength 532 nm), spatial filter, collimating lenses, and high-speed camera 2 with a Tokina lens (focal length of $100$ mm and maximum aperture $f/2.8$, model: AT-X M100 PRO D Macro) positioned at $x=180$ mm as shown in figure \ref{fig1} are used for digital in-line holography. The spatial filter consisting of an infinity-corrected plan achromatic objective (20X magnification, make: Holmarc Opto-Mechatronics Ltd.) and a 15 $\mu$m pin-hole is used to produce a clean beam that is expanded using a plano-concave lens and then collimated using a plano-convex lens that illuminates the droplet field of view. The resultant interference patterns created due to the child droplets are recorded using the high-speed camera 2 with a resolution of $2048 \times 1600$ pixels at 1800 fps with an exposure duration of 1 $\mu$s and spatial resolution of 15.56 $\mu$m/pixel. In the following section, the reported size distributions are obtained from three repetitions for each set of parameters. Figure \ref{figS2} outlines the various steps involved in digital in-line holography processing. A detailed illustration of the experimental setup, shadowgraphy, digital in-line holography, and the associated post-processing method can also be found in \cite{ade2022droplet}. 

\section{Results and discussion}
\label{sec:dis}
\begin{figure}
\centering
\includegraphics[width=0.9\textwidth]{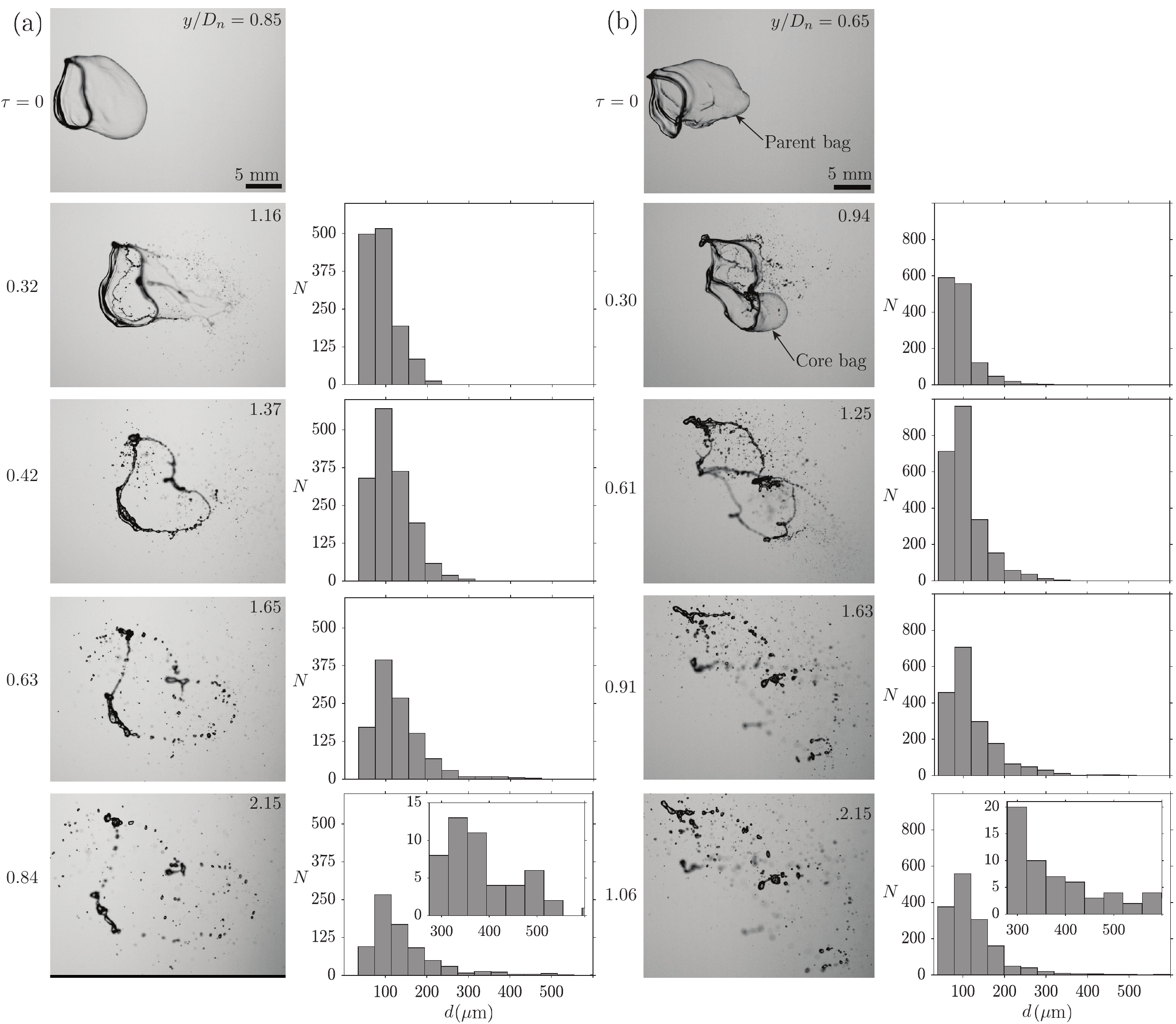}
\caption{Temporal evolution of the fragmentation process (from shadowgraphy) and droplet size distribution (the droplet counts, $N$ versus the droplet diameter, $d$) for (a) $\We = 12.6$ and (b) $\We = 34.8$. The dimensionless time and droplet position in the flow direction $(y/D_{n})$ are mentioned in each panel. Here, $\tau=0$ represents the onset of breakup. The droplet breakup phenomena for $\We = 12.6$ and 34.8 are provided as supplementary movies 2 and 3, respectively. The corresponding time-series focused holograms are included as supplementary movies 4 and 5, respectively.}
\label{fig2}
\end{figure}

A drop in an airstream exhibits different morphologies and breakup modes as the Weber number increases \citep{soni2020deformation,kirar2022experimental,boggavarapu2021secondary}. For a low Weber number, the drop enters the potential core of the airstream and deforms into a disk due to aerodynamic force while the surface tension acts to bring it to a spherical shape. Thus, the drop undergoes shape oscillations and fragments due to capillary instability and generates child droplets of comparable sizes (mono-modal size distribution). This is termed as transitional breakup (see, supplementary movie 1 for $\We = 11.4$). As the aerodynamic force increases, the drop forms a single bag for $\We=12.6$  (figure \ref{fig2}a) and dual-bag for $\We=34.8$ (figure \ref{fig2}b). Subsequently, the bags, rim, and nodes undergo fragmentation due to Rayleigh–Taylor instability, Rayleigh–Plateau capillary instability  \citep{taylor1963shape} and nonlinear instability at different stages \citep{jackiw2021aerodynamic,jackiw2022prediction,kirar2022experimental}. Supplementary movies 2 and 3 show the breakup phenomena for $\We=12.6$ and $34.8$, respectively. In the following, we present the temporal evolution of the DSD resulting from dual-bag fragmentation ($\We=34.8$) and contrast the dynamics with that associated with single-bag fragmentation ($\We=12.6$).

Figure \ref{fig2}(a) and (b) illustrates the temporal evolution of the DSD at $\We = 12.6$ (single-bag fragmentation) and $\We=34.8$ (dual-bag fragmentation), respectively. In figure \ref{fig2}(a), it can be seen that the drop deforms to an elongated bag with a toroidal rim shape due to the aerodynamic force at the onset of rupture of the bag ($\tau=0$). As a result of this rupture, tiny child droplets are produced ($d < 300 \hspace{1mm} \mu$m at $\tau= 0.32$). Subsequently, the remaining portion of the bag breaks and produces tiny child droplets at $\tau= 0.42$. At this stage, the fragmentation of the toroidal rim due to capillary instability initiates and continues till $\tau=0.63$. This rim breakup produces intermediate size ($300 \hspace{1mm} \mu{\rm m} < d < 600 \hspace{1mm} \mu{\rm m}$) droplets, as evident in the corresponding histogram at $\tau=0.63$. Finally, the nodes detach from the rim and fragment due to Rayleigh-Taylor instability and generate bigger droplets at $\tau=0.84$. The inset in the histogram at $\tau=0.84$ shows the zoomed view for $300 \hspace{1mm} \mu{\rm m} < d < 600 \hspace{1mm} \mu{\rm m}$.

In contrast, in figure \ref{fig2}(b), the inflated parent bag (at the onset of breakup) surrounded by parent rim and undeformed core is shown at $\tau=0$. At $\tau=0.30$, tiny child droplets of diameter with $d<300 \hspace{1mm} \mu$m are produced due to the rupture of the parent bag under the aerodynamic action. At this stage, the aerodynamic field constantly imparts the body force on the undeformed core drop, hence making the local Weber number sufficiently large for the second breakup process. Therefore, the core bag (second bag) is drawn out from the core drop's edge after the parent bag rupture. In the next stage ($\tau=0.61$), the core bag bursts and forms tiny child droplets of diameter $d<300 \hspace{1mm} \mu$m. As a result, the total number of tiny child droplets increases at $\tau=0.61$. Further, the rims associated with the initial main drop and smaller core drop fragments due to capillary instability generate intermediate-sized droplets ($300 \hspace{1mm} \mu{\rm m} < d < 600 \hspace{1mm} \mu{\rm m}$) at $\tau=0.91$. At $\tau = 1.06$, the nodes associated with the rims of the parent and core drops detach/break. This process generates bigger-sized child droplets of diameter $d>600 \hspace{1mm} \mu$m (not shown). The inset in the histogram at $\tau=1.06$ depicts the zoomed view in the region $300 \hspace{1mm} \mu{\rm m} < d < 600 \hspace{1mm} \mu{\rm m}$. The number of tiny child droplets decreases later as they move out of the field of view because of their high velocities. Secondly, there may be some coalescence of small droplets during the fragmentation process, leading to an increase in the number of larger child droplets at later times.

\begin{figure}
\centering
\hspace{0.8cm}(a) \hspace{5.5cm}(b) \\
\includegraphics[height=0.3\textwidth]{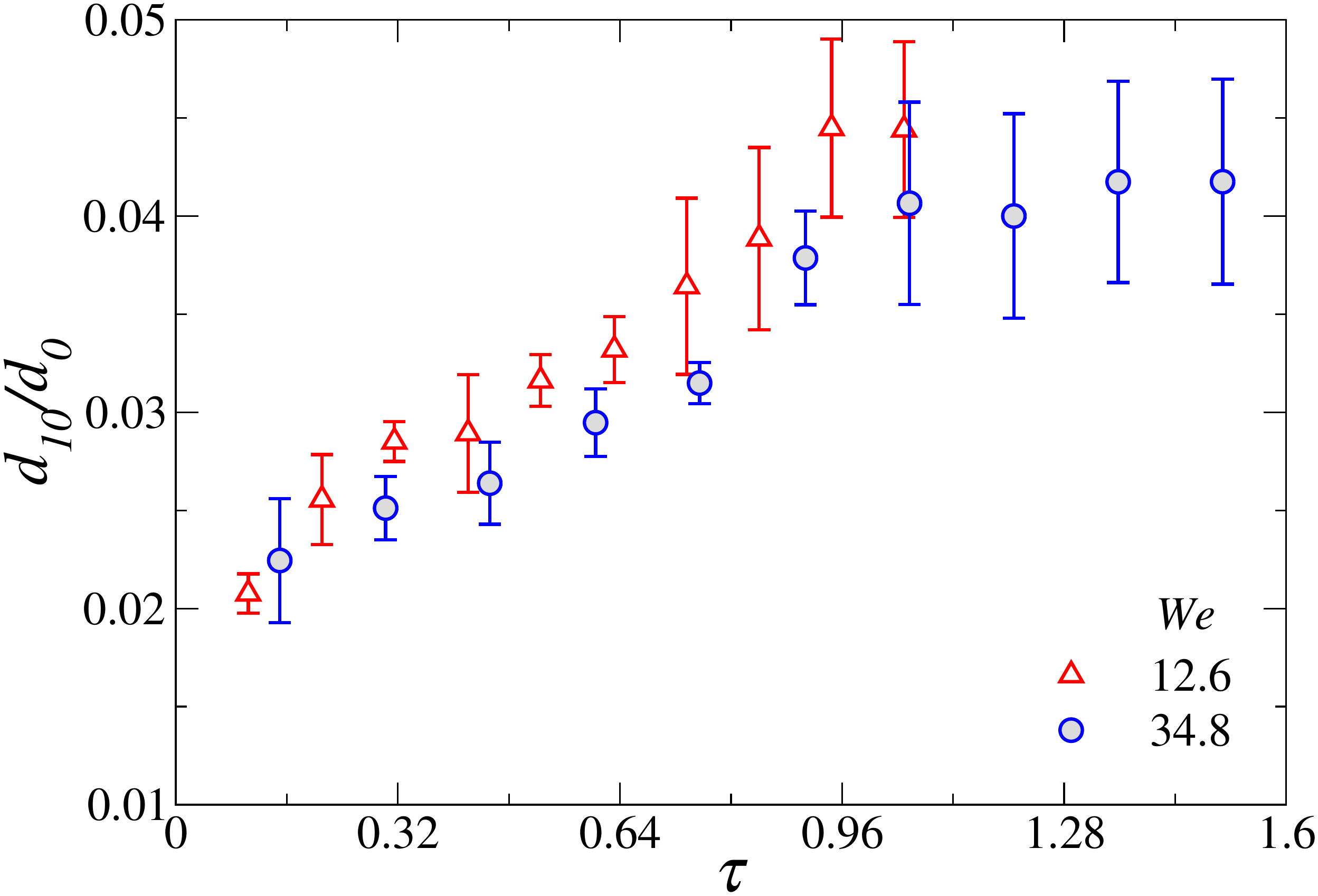} \ \includegraphics[height=0.3\textwidth]{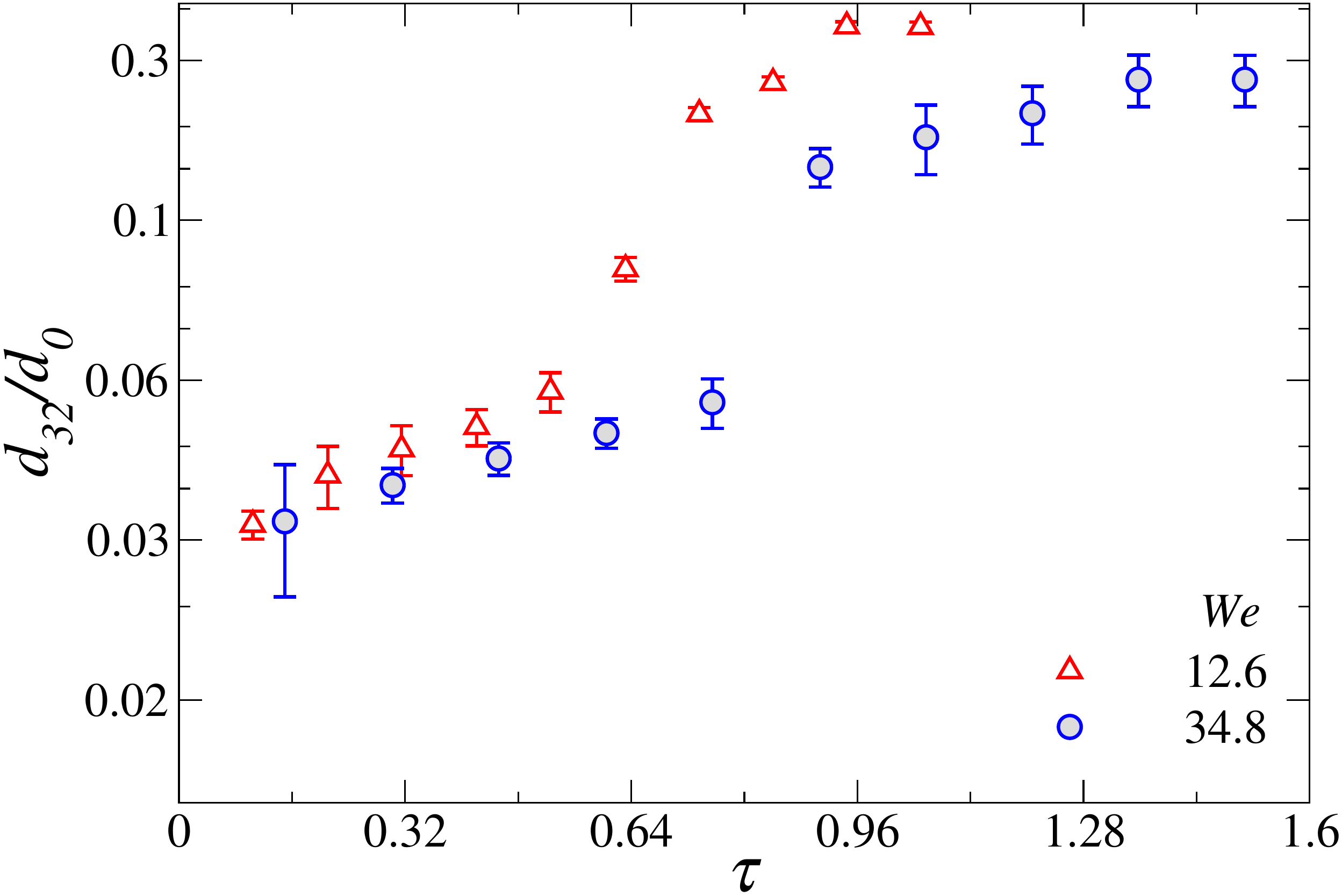}
\caption{Temporal variation of the normalised (a) number mean diameter $(d_{10}/d_0)$ and (b) Sauter mean diameter $(d_{32}/d_0)$ for $\We=12.6$ and 34.8. The error bar represents the standard deviation obtained using three repetitions.}
\label{fig3}
\end{figure}

The temporal variations of the normalised number mean diameter $(d_{10}/d_0=\int_{0}^{\infty}dp(d)\textrm{d}d/d_0)$ and Sauter mean diameter $(d_{32}/d_0 = {\int_{0}^{\infty}d^{3}p(d)\textrm{d}d / \int_{0}^{\infty }d^{2}p(d)\textrm{d}d}/d_0)$ for $\We=12.6$ and $\We=34.8$ are plotted in figure \ref{fig3}(a) and (b), respectively. Here, $p(d)$ is the probability density function of $d$. It can be seen in figure \ref{fig3}(a) that $d_{10}/d_{0}$ increases with time for both single-bag and dual-bag breakup cases. This is because, while the bag rupture initially only creates small child droplets, later on during the fragmentation process, larger child droplets are created, primarily generated by the fragmentation of the rim and nodes. The reciprocal of Sauter mean diameter, $1/d_{32}$ signifies the measure of surface area per unit volume of child droplets. It can be seen that the temporal variations of $d_{32}/d_0$ for single-bag and dual-bag breakups mostly overlap at the early stage $(\tau \le 0.3)$, but at later times $d_{32}/d_0$ for the dual-bag breakup is significantly lower than that of the single-bag breakup. This indicates that the fragmentation of the second bag at the later stage contributes to the number of tiny child droplets. It can be observed in figure \ref{fig3}(a) and (b) that the variations of $d_{10}$ and $d_{32}$ reaches a plateau at $\tau \approx 0.84$, 1.06 for $\We=12.6$, 34.8, respectively. This behaviour indicates that the fragmentation approximately ceases at this instant.

\subsection{Prediction of overall size distribution} \label{sec31}
The volume probability $(P_v)$ is the ratio of the total volume of droplets of a specific diameter to the total volume of all droplets. This is given by \citep{jackiw2022prediction} 
\begin{equation} \label{j:eq1}
P_{v}=\frac{\zeta ^{3}P_{n}}{\int_{0}^{\infty}\zeta ^{3}P_{n}d\zeta}={\frac{\zeta ^{3}P_{n}}{\beta^{3}\Gamma (\alpha+3)/\Gamma (\alpha)}},
\end{equation}
wherein $P_n= {\zeta^{\alpha -1}e^{-\zeta /\beta} / \beta^{\alpha}\Gamma (\alpha)}$; $\zeta \left (=d/d_0 \right)$; $\Gamma (\alpha)$ represents the gamma function; $\alpha=(\bar{\zeta}/\sigma_s)^{2}$ and $\beta=\sigma_s^{2}/\bar{\zeta}$ are the shape and rate parameters, respectively; $\bar{\zeta}$ and $\sigma_s$ are the mean and standard deviation of the distribution, which are estimated based on the characteristic breakup sizes corresponding to each mode. 

In the single-bag breakup, the total size distribution results from three breakup modes, namely bag, rim, and node breakups. Thus, the overall volume probability density $P_{v,Total}$ is a weighted sum of each mode. \cite{jackiw2022prediction}  developed an analytical model and used it to predict the overall volume probability density and characteristic sizes of the child droplets for the single-bag and sheet-thinning breakups. In the present study, we perform experiments to investigate the droplet morphology for different Weber numbers and utilize the \cite{jackiw2022prediction} model to estimate the overall volume probability density and characteristic sizes of the child droplets resulting from single, dual, and multi-bag breakups.

In the dual-bag fragmentation, the overall size distribution ($P_{v,Total}$) is the combination of DSD associated with the parent ($P_{v,p}$) and core ($P_{v,c}$) drops as shown in figure \ref{fig2}(b). Thus,
\begin{equation} \label{j:eq3}
P_{v,Total}=P_{v,p}+P_{v,c},
\end{equation}
where the contribution of the parent  drop is given by
\begin{equation} \label{j:eq4}
P_{v,p}=w_{N}P_{v,N}+w_{R}P_{v,R}+w_{B}P_{v,B},
\end{equation}
where $w_{N}=V_{N}/V_{0}$, $w_{R}=V_{R}/V_{0}$ and $w_{B}=V_{B}/V_{0}$ represent the contributions of the volume weights from the node, rim, and bag of the parent drop, respectively. Here, $V_{0}$, $V_{B}$, $V_{R}$, $V_{N}$ denote the volumes of the initial drop, bag, rim, and node, respectively.

Due to the aerodynamic force, at the early stage, the spherical drop deforms into a disk, which in turn inflates into a bag. The volume of the deformed parent drop, $V_D$ is given by \citep{jackiw2021aerodynamic}
\begin{equation} \label{j:eq5}
\frac{V_{D}}{V_{0}}=\frac{3}{2}\left [ \left ( \frac{2R_{i}}{d_{0}} \right )^{2}\left ( \frac{h_{i}}{d_{0}} \right )-2\left ( 1-\frac{\pi }{4} \right )\left ( \frac{2R_{i}}{d_{0}} \right )\left ( \frac{h_{i}}{d_{0}} \right )^{2} \right ],
\end{equation}
where $h_i$ denotes the disk thickness and $2R_i$ represents the major diameter of the rim, which can be evaluated as \citep{jackiw2021aerodynamic,jackiw2022prediction}
\begin{equation} \label{j:eq7}
\frac{h_{i}}{d_{0}}=\frac{4}{\We_{rim}+10.4}, \hspace{2mm} {\rm and} \hspace{2mm}
\frac{2R_{i}}{d_{0}}=1.63-2.88e^{({-0.312\We})}.
\end{equation}
Here, $\We_{rim} (=\rho_w \dot{R}^{2}{d_{0}}/\sigma)$ denotes the rim Weber number, which indicates the competition between the radial momentum induced at the drop periphery and the restoring surface tension of the stable drop. The constant radial expansion rate of the drop, $\dot{R}$ is given by $\dot{R}=(1.125 U \sqrt{\rho_a/\rho_w)}/2)(1-32/9 \We)$ \citep{jackiw2022prediction}. Using these expressions, we can estimate $w_{N}$, $w_{R}$ and $w_B$ as 
\begin{eqnarray} 
w_{N} &=&\frac{V_{N}}{V_{0}}=\frac{V_{N}}{V_{D}}\frac{V_{D}}{V_{0}}, \label{j:eq6} \\ 
w_{R} &=&\frac{V_{R}}{V_{0}}=\frac{3\pi }{2}\left [ \left ( \frac{2R_{i}}{d_{0}} \right )\left ( \frac{h_{i}}{d_{0}} \right )^{2}-\left ( \frac{h_{i}}{d_{0}} \right )^{3} \right ],\label{j:eq6b} \\
w_B &=&\frac{V_{B}}{V_{0}} =\frac{V_{D}}{V_{0}}-\frac{V_{N}}{V_{0}}-\frac{V_{R}}{V_{0}},  \label{j:eq9}
\end{eqnarray}
where $V_N/V_D$ represents the volume fraction of the node relative to the disk. \cite{jackiw2022prediction} experimentally investigated the bag and bag-stamen breakups at different Weber numbers and estimated the mean value of $V_N/V_D$ to be 0.4 for bag breakups. Thus, in our study, we use $V_N/V_D=0.4$ to estimate $w_N$.

The characteristic sizes associated with the node, rim, and bag are then separately estimated for the parent and core drops. \cite{jackiw2022prediction} used a similar analysis, albeit for the single-bag and sheet-thinning fragmentation process.\\

\textit{Parent node:} The node breakup occurs due to the Rayleigh-Taylor (RT) and Rayleigh-Plateau instabilities, respectively \citep{zhao2010morphological,kirar2022experimental}. The child droplet size associated with node ($d_{N}$) breakup for the parent drop is given by \citep{jackiw2022prediction}
\begin{equation} \label{j:eq10}
{d_{N}}= d_{0}\left [ \frac{3}{2}\left ( \frac{h_{i}}{d_{0}} \right )^{2}\frac{\lambda_{RT} }{d_{0}}n \right ]^{1/3}.
\end{equation}
Here, $\lambda_{RT}=2\pi\sqrt{3\sigma/\rho_w a}$ is the maximum susceptible wavelength of the RT instability, wherein $a=\frac{3}{4}C_{D}\frac{U^{2}}{d_{0}}\frac{\rho_{a} }{\rho_{w}}\left ({D_{max}/d_{0}} \right )^{2}$ is the acceleration of the deforming droplet. The drag coefficient ($C_{D}$) of the disk shape droplet is about 1.2 and the extent of droplet deformation is given by ${D_{max} / d_0}={2 / (1+\exp{(-0.0019 \We^{2.7})})}$ \citep{zhao2010morphological}. In Eq. (\ref{j:eq10}), $n=V_{N}/V_{D}$ indicates the volume fraction of the nodes relative to the disk. \cite{jackiw2022prediction} estimated the minimum, mean and maximum values of $n$ to be 0.2, 0.4, and 1, respectively. The node droplets exhibit three characteristic sizes ($d_N$), which can be determined using these three values of $n$. The mean and standard deviation of the parent node distribution are calculated from these three characteristic sizes. \\

\textit{Parent rim:} There are three mechanisms for rim breakup. They are (i) the Rayleigh-Plateau instability, (ii) the receding rim instability, and (iii) the nonlinear instability of liquid ligaments near the pinch-off point. The child droplet size $(d_{R})$ produced from the first mechanism is given by \citep{jackiw2021aerodynamic}
\begin{equation} \label{j:eq11}
d_{R}=1.89h_{f},
\end{equation}
where $h_{f}=h_{i}\sqrt{R_{i}/R_{f}}$ is the final rim thickness. The term $R_f$ in the above expression represents the radius of the bag at the time of its burst and it can be evaluated as \citep{kirar2022experimental}
\begin{equation} \label{new_eq1}
R_f=\frac{d_{0}}{2\eta} \left [ 2e^{\tau ^{\prime}\sqrt{p}}+\left ( \frac{\sqrt{p}}{\sqrt{q}}-1 \right )e^{-\tau ^{\prime}\sqrt{q}}-\left ( \frac{\sqrt{p}}{\sqrt{q}}+1 \right )e^{\tau ^{\prime}\sqrt{q}} \right ],
\end{equation}
where $\eta = f^{2}-120/\We$, $p=f^{2}-96/\We$ and $q=24/\We$. The value of the stretching factor, $f$ for a droplet undergoing breakup in a cross-flow of air is $2\sqrt{2}$ \citep{kulkarni2014bag}. In Eq. (\ref{new_eq1}), the dimensionless time, $\tau ^{\prime}=Ut_{b}\sqrt{\rho_{a}/\rho _{w} }/d_{0}$. Here, the bursting time, $t_b$, can be evaluated as \citep{jackiw2022prediction}
\begin{equation} \label{new_eq2}
t_{b}=\frac{\left [ \left ( \frac{d_{i}}{d_{0}} \right )-2\left ( \frac{h_{i}}{d_{0}} \right ) \right ]}{\frac{2\dot{R}}{d_{0}}}\left [ -1+\sqrt{1+9.4 \frac{8t_{d}}{\sqrt{3\We}}\frac{\frac{2\dot{R}}{d_{0}}}{\left [ \left ( \frac{d_{i}}{d_{0}} \right )-2\left ( \frac{h_{i}}{d_{0}} \right ) \right ]}\sqrt{\frac{V_{B}}{V_{0}}}} \right ],
\end{equation}
where $d_{i}=2 R_i$ and $t_d=d_0/U\sqrt{\rho_w/\rho_a}$ denotes the deformation time scale.

The child droplet size ($d_{rr}$) associated with the second mechanism is given by \citep{jackiw2022prediction}
\begin{equation} \label{j:eq12}
d_{rr}=d_{0}\left [ \frac{3}{2}\left ( \frac{h_{f}}{d_{0}} \right )^{2}\frac{\lambda_{rr}}{d_{0}} \right ]^{1/3},
\end{equation}
where $\lambda _{rr}=4.5b_{rr}$ is the wavelength of the receding rim instability, $b_{rr}=\sqrt{\sigma /\rho_w a_{rr}}$ is the receding rim thickness. Here, $a_{rr}=U_{rr}^{2}/R_{f}$ denotes the acceleration of the receding rim \citep{wang2018universal}, wherein $U_{rr}$ represents the receding rim velocity. After the bag ruptures, the edge begins to roll up, forming a receding rim that recedes along the bag. In the present study, $U_{rr}$ is calculated experimentally by measuring the rate of displacement of the receding rim relative to the main rim.

The third mechanism produces two different characteristic sizes of child droplets, which are given by \citep{keshavarz2020rotary}
\begin{equation} \label{j:eq13}
d_{sat,R}=\frac{d_R}{\sqrt{2+3Oh_{R}/\sqrt{2}}} ~~ \textrm{and}~~
d_{sat,{rr}}=\frac{d_{rr}}{\sqrt{2+3Oh_{R}/\sqrt{2}}},
\end{equation}
where $Oh_{R} (=\mu /\sqrt{\rho_w h_{f}^{3}\sigma})$ denotes the Ohnesorge number based on the final parent rim thickness. The characteristic sizes are given by Eqs. (\ref{j:eq11}), (\ref{j:eq12}) and (\ref{j:eq13}) are used to evaluate the number-based mean and standard deviation for the parent rim distribution.\\

\textit{Parent bag:} The characteristic sizes associated with the bag fragmentation due to the minimum bag thickness ($d_{B}$), the receding rim thickness ($d_{rr}$), the Rayleigh-Plateau instability ($d_{RP,B}$), and nonlinear instability of liquid ligaments ($d_{sat,B}$) are given by \citep{jackiw2022prediction}
\begin{equation} \label{j:eq15}
d_{B}=h_{min}, ~  d_{rr,B}=b_{rr}, ~
d_{RP,B}=1.89 b_{rr}, ~ \textrm{and}~ d_{sat,B}=\frac{d_{RP,B}}{\sqrt{2+3Oh_{rr}/\sqrt{2}}},
\end{equation}  
where $h_{min}= (2.3\pm1.2)$ $\mu$m is the minimum bag thickness as reported by \cite{jackiw2022prediction} and $Oh_{rr} (=\mu /\sqrt{\rho_{w} b_{rr}^{3}\sigma})$ is the Ohnesorge number based on receding rim thickness. These characteristic sizes are used to estimate the number mean and standard deviation associated with the parent bag fragmentation mode. \\

\textit{Core droplet size distribution ($P_{v,c}$):} The core drop also exhibits three modes similar to the parent drop. The weight of each breakup mode should be multiplied by the volume weight of the core drop as it only comprises up a small portion of the original parent drop. Thus, by taking into account the volume weight of each mode, the combined size distribution of the core drop $(P_{v,c})$ is given by \citep{jackiw2022prediction}
\begin{equation} \label{j:eq16}
P_{v,c}=\frac{V_{c}}{V_{0}}(w_{N,c}P_{v,c,N}+w_{R,c}P_{v,c,R}+w_{B,c}P_{v,c,B}),
\end{equation}
where, $w_{B,c}=V_{Bc}/V_{c}$, $w_{R,c}=V_{Rc}/V_{c}$ and $w_{N,c}=V_{Nc}/V_{c}$ denote the contributions of volume weights from the core bag, core rim, and core node, respectively. Here, $V_{Bc}$, $V_{Rc}$, and $V_{Nc}$ are the bag,  rim, and node volumes of the core drop, respectively. The volume of the core drop, $V_c$ is given by $V_{c}=V_{0} (1-{V_D/V_0})$. The volume weight of each breakup mode for the core drop is estimated in the same manner as the parent drop. However, the Weber number of the core drop is determined by considering the relative velocity between airflow and core drop as explained in \cite{jackiw2022prediction}. In addition, the characteristic sizes corresponding to each breakup mode of the core drop are determined in the same manner as described for the parent drop. Note that the characteristic sizes for the core drop are represented using a subscript `$c$'. 

\begin{figure}
\centering
\includegraphics[width=0.95\textwidth]{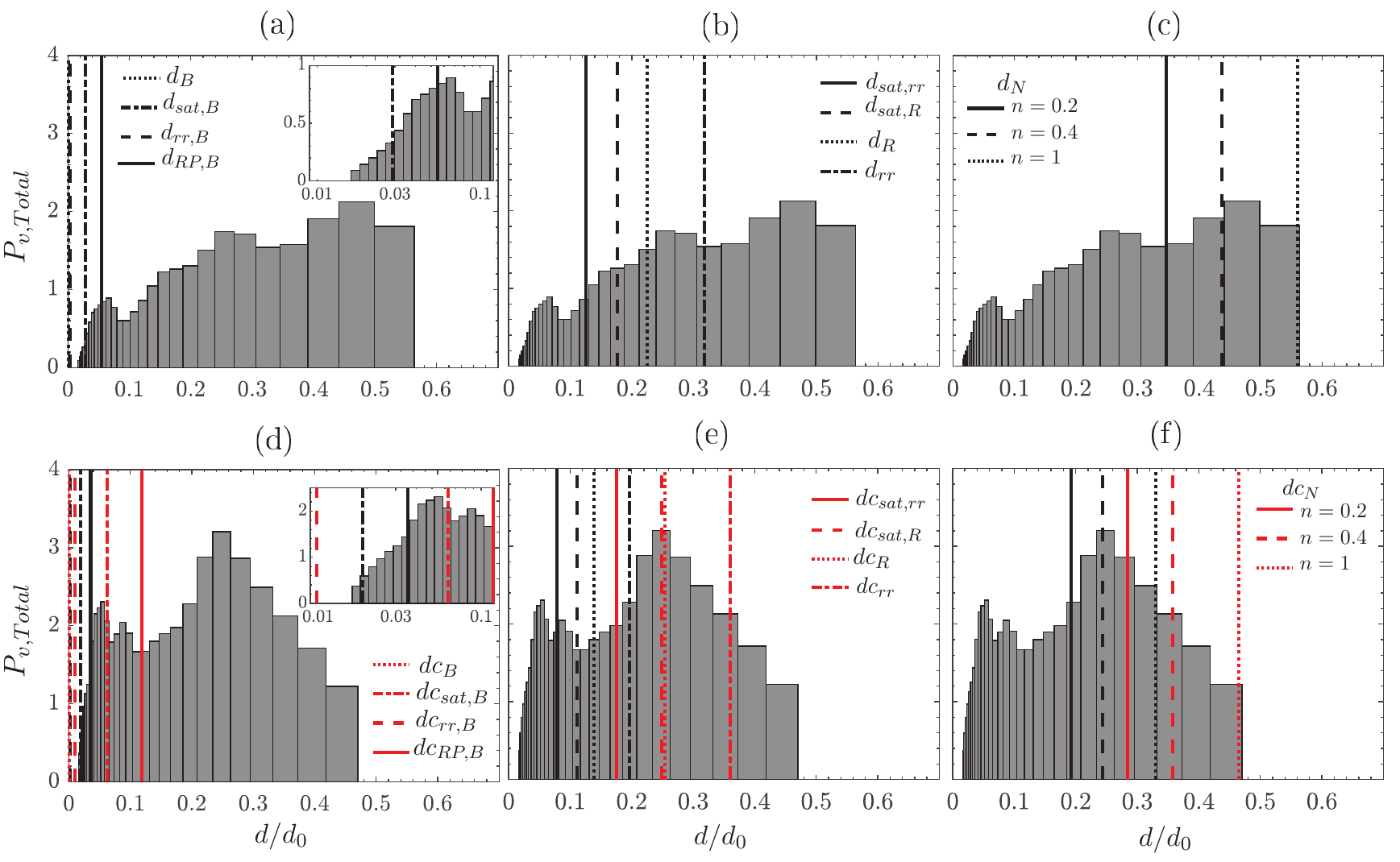}
\caption{Comparison of the theoretically predicted droplet characteristic breakup sizes with experimental data: (a,b,c) $\We = 12.6$ at $\tau=0.84$ and (d,e,f) $\We=34.8$ at $\tau=1.06$. Here, the black and red lines represent the characteristic sizes of the parent and core drops, respectively. The insets in panels (a) and (c) show the enlarged views for small-size child droplets with the axis for $d/d_0$ in $\log-$scale.}
\label{fig4}
\end{figure}

\begin{figure}
\centering
\includegraphics[width=0.9\textwidth]{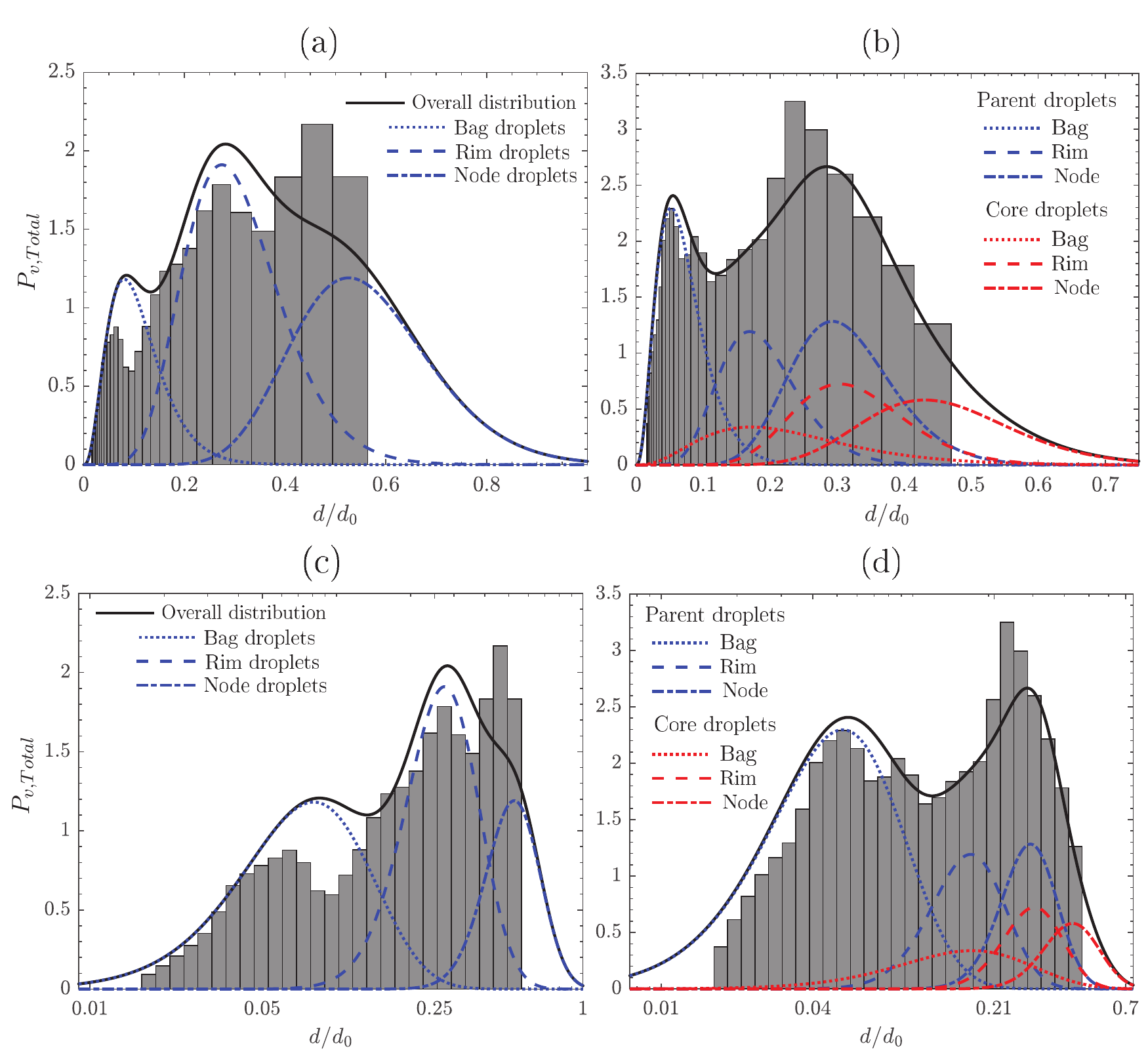}
\caption{Comparison of the multi-modal distribution obtained from the experiments with the analytical predictions for (a) $\We = 12.6$ at $\tau=0.84$ (single-bag breakup) and (b) $\We = 34.8$ at $\tau=1.06$ (dual-bag breakup). The solid line represents the size distribution due to the sum of all modes (overall size distribution). Panels (c) and (d) present the results of panels (a) and (b) in $\log-$ scale for $d/d_0$ axis.}
\label{fig5}
\end{figure}

Figure \ref{fig4}(a-c) and \ref{fig4}(d-f) depict the comparison of the theoretically predicted characteristic breakup sizes of a drop undergoing single-bag and dual-bag fragmentation processes with experimental results. The panels \ref{fig4}(a,d), \ref{fig4}(b,e), and \ref{fig4}(c,f) are for the bag, rim, and node breakup modes, respectively. In contrast to node fragmentation, which results in three characteristic sizes, the bag, and rim breakups produce four characteristic sizes of the child droplets, as discussed in \cite{jackiw2022prediction} for a single-bag breakup. It can be seen in figure \ref{fig4}(a-c) and \ref{fig4}(d-f) that for both single-bag and dual-bag breakups, the theoretically predicted sizes agree well with the experimental results. In other words, the peaks of the experimental distribution correspond to the mean values of the characteristic sizes for individual modes. In the dual-bag breakup (figure \ref{fig4}d-f), the characteristic sizes produced from the rim, bag, and node of the parent drop are smaller than the corresponding fragmentation of the core drop. This is because the parent drop experiences a strong aerodynamic field while the core drop encounters a weaker aerodynamic field as it moves away from the nozzle. Therefore, the local Weber number corresponding to the core drop ($\We_c \approx 17$) is significantly lower than that of the parent drop ($\We \approx 34.8$). These predicted characteristic sizes are used to evaluate the individual size distribution of each mode (bag, rim, and node), and then the total volume probability distribution is estimated. 

The comparison of the size distribution obtained from the analytical model and experiments are shown in figure \ref{fig5}(a) and \ref{fig5}(b) for the single-bag and dual-bag breakups, respectively. In the case of a single-bag breakup, it can be seen in figure \ref{fig5}(a) that the contributions from the bag and rim are slightly over-predicted, while it is under-predicted for the node. In figure \ref{fig5}(a), the first ($d/d_0\approx 0.06$), second ($d/d_0\approx 0.27$), and third ($d/d_0\approx 0.46$) peaks are associated with the bag, rim and nodes, respectively indicating a tri-modal distribution, as also reported by \cite{guildenbecher2017characterization,jackiw2022prediction}. In contrast to the single-bag, dual-bag fragmentation involves the breakup of parent and core drops distinctly. Each drop undergoes three breakup modes associated with the bag, rim, and nodes. Thus, six physical processes contribute to the overall size distribution. The experimental results in figure \ref{fig5}(b) show that the first ($d/d_{0}\approx 0.05$) and second peaks ($d/d_{0}\approx 0.09$) are close to each other and are likely due to the fragmentation of parent and core bags, respectively. The parent bag experiences a strong aerodynamic field and thus undergoes more inflation leading to smaller child droplets. On the other hand, the core drop is exposed to a weaker aerodynamic field and thus elongates less, which results in bigger child droplets. The third peak ($d/d_{0}\approx 0.25$) in figure \ref{fig5}(b) is due to the fragmentation of the rim and nodes of the parent and core drops. Despite the complex breakup phenomenon involving six distinct modes, the resultant overall size distribution exhibits a bi-model size distribution in the dual-bag breakup. This can be explained by analysing figure \ref{fig4}(d-f) and figure \ref{fig5}(b), which depict that the size of the child droplets from the core bag overlaps with that of the parent rim. The size of child droplets resulting from the core rim and parent node breakup also lies in the same range. These observations are distinct from that of the single-bag fragmentation. It is to be noted that \cite{jackiw2022prediction} analytically showed that while single-bag breakup exhibits a tri-modal distribution, the sheet-thinning breakup displays a mono-modal distribution.

\subsection{Contribution from individual breakup modes}

\begin{figure}
\centering
\includegraphics[width=0.8\textwidth]{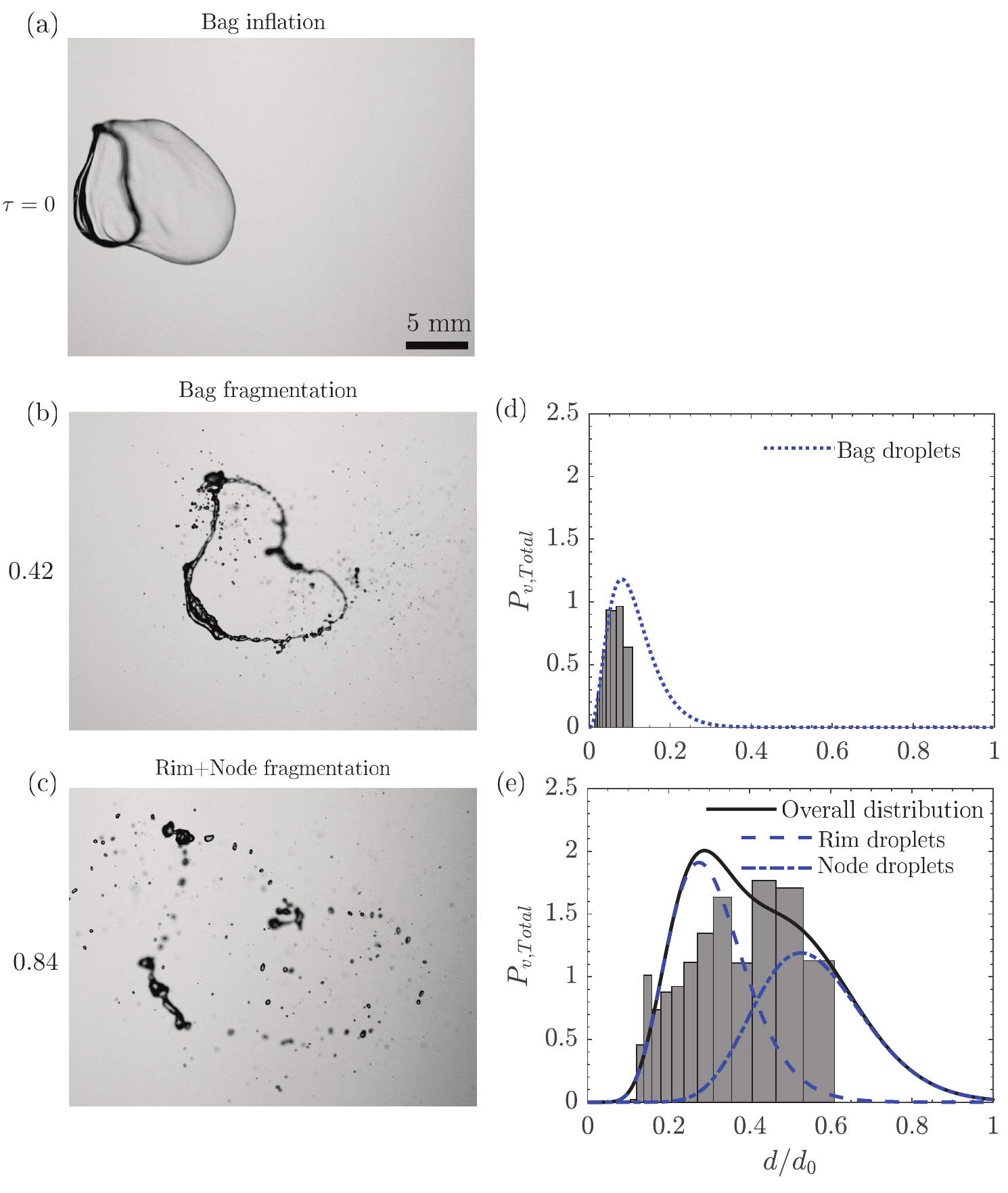}
\caption{Mode decomposition for single-bag breakup at $We=12.6$. Panels (a-c) depict shadowgraphy images at three instants corresponding to the maximum bag inflation $(\tau=0)$, rupture of bag $(\tau=0.42)$, and fragmenting of the rim and node $(\tau=0.84)$, respectively. The corresponding size distributions of child droplets at $\tau=0.42$ and 0.84 are shown in panels (d) and (e), respectively.}
\label{fig6}
\end{figure}

\begin{figure}
\centering
\includegraphics[width=0.90\textwidth]{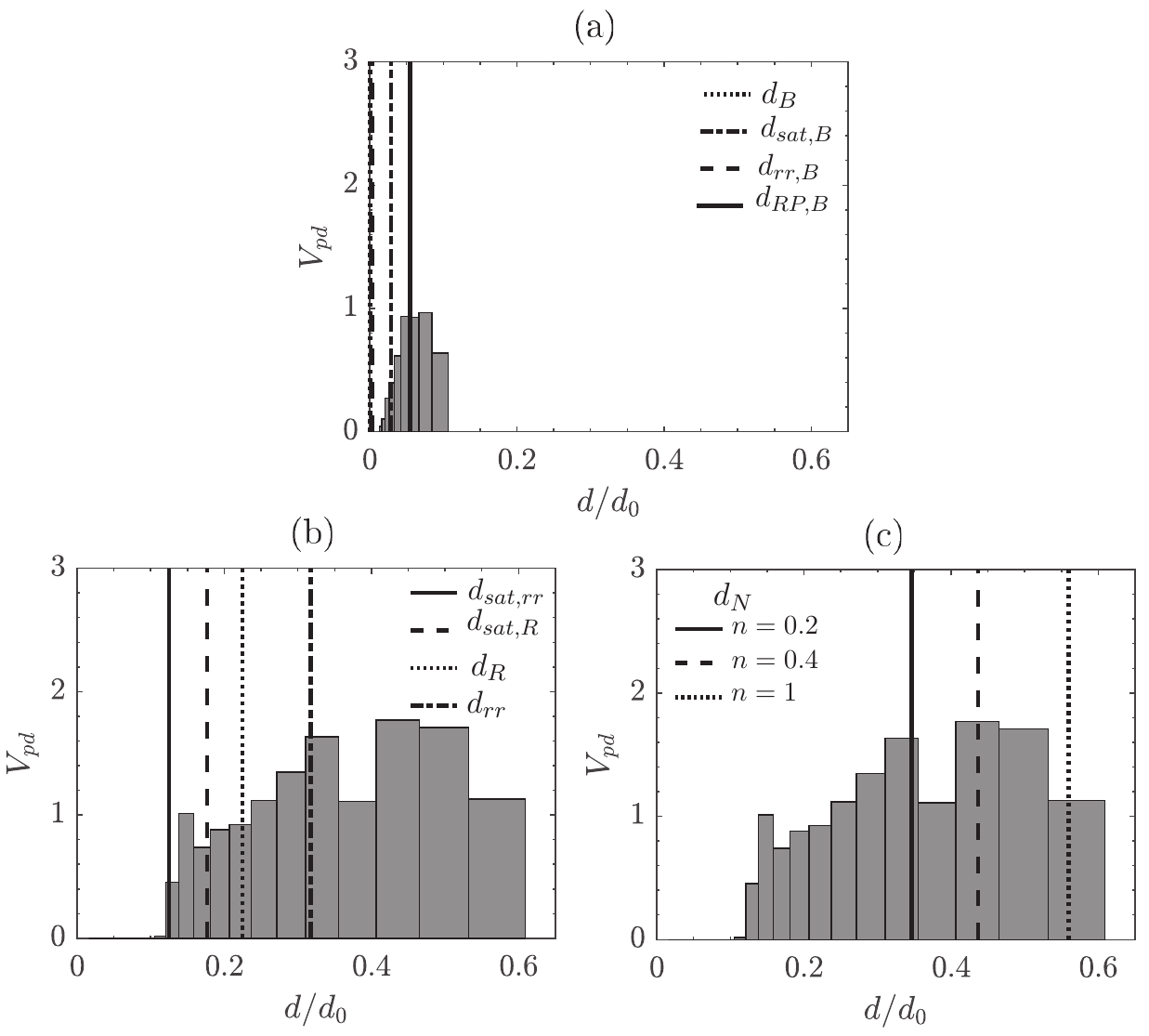}
\caption{Comparison of experimental data with the theoretically predicted characteristic breakup sizes at (a) $\tau=0.42$, and (b,c) $\tau=0.84$ for $\We = 12.6$ (single-bag breakup). The panels (b) and (c) correspond to the rim and node fragmentations, respectively.}
\label{Appendix_fig1}
\end{figure}

To get better insight into the breakup phenomenon, we perform mode decomposition to evaluate the individual contributions of different modes to the overall size distribution of the child droplets. The mode decomposition for the single-bag breakup ($\We=12.6$) is presented in figure \ref{fig6}. At $t=0.42$, the bag completely fragments, as shown in figure \ref{fig6}(b). The resulting size distribution is shown in figure \ref{fig6}(d). In the analytical model, we use the contribution of the bag only as given in \cite{jackiw2022prediction}. It is evident that the analytical model predicts the experimentally observed size distribution for the bag fragmentation that depicts a mono-modal size distribution. In order to isolate the contributions of the rim and nodes, we subtract the child droplets generated by the bag fragmentation from the total fragments at $\tau=0.84$ (figure \ref{fig6}e). However, since the rim and nodes break simultaneously, it is difficult to distinguish their individual contributions. It can be seen that both experimental and analytical distributions exhibit a bi-modal distribution. Inspection of figures \ref{fig6}(d) and (e) reveals that the distributions for bag, rim, and nodes exhibit distinct peaks associated with different values of $d/d_0$. As a result, the combined contributions of the bag, rim, and nodes result in a tri-modal distribution, as shown in figure \ref{fig5}(a). Quantitatively, the total volume percentage of the child droplets due to bag, rim+node are 14.88\% and 85.12\%, respectively. The characteristic sizes at the instants associated with the rupture of the bag $(\tau=0.42)$ and the fragmentation of the rim and nodes $(\tau=0.84)$ are presented in figure \ref{Appendix_fig1}(a) and (b), respectively.

\begin{figure}
\centering
\includegraphics[width=0.8\textwidth]{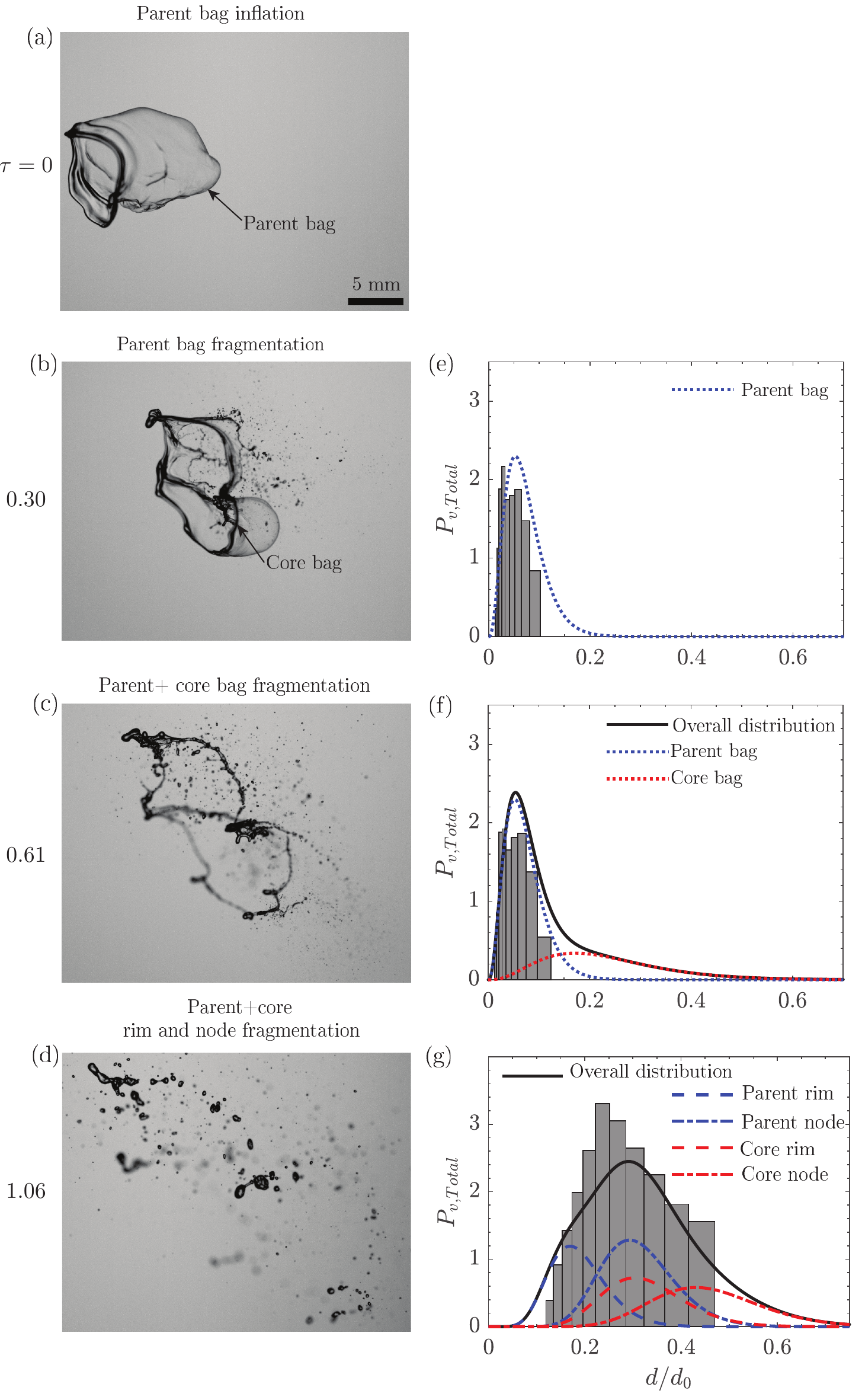}
\caption{Mode decomposition for the dual-bag breakup at $We=34.8$. Panels (a-d) depict shadowgraphy images at four instants corresponding to the maximum inflation of the parent bag $(\tau=0)$, rupture of parent bag $(\tau=0.3)$, fragmenting of core bag $(\tau=0.61)$ and fragmenting of rim + nodes $(\tau=1.06)$, respectively. The corresponding size distributions of child droplets at $\tau=0.3$, 0.61, and 1.06 are shown in panels (e), (f), and (g), respectively.}
\label{fig7}
\end{figure}

\begin{figure}
\centering
\includegraphics[width=0.90\textwidth]{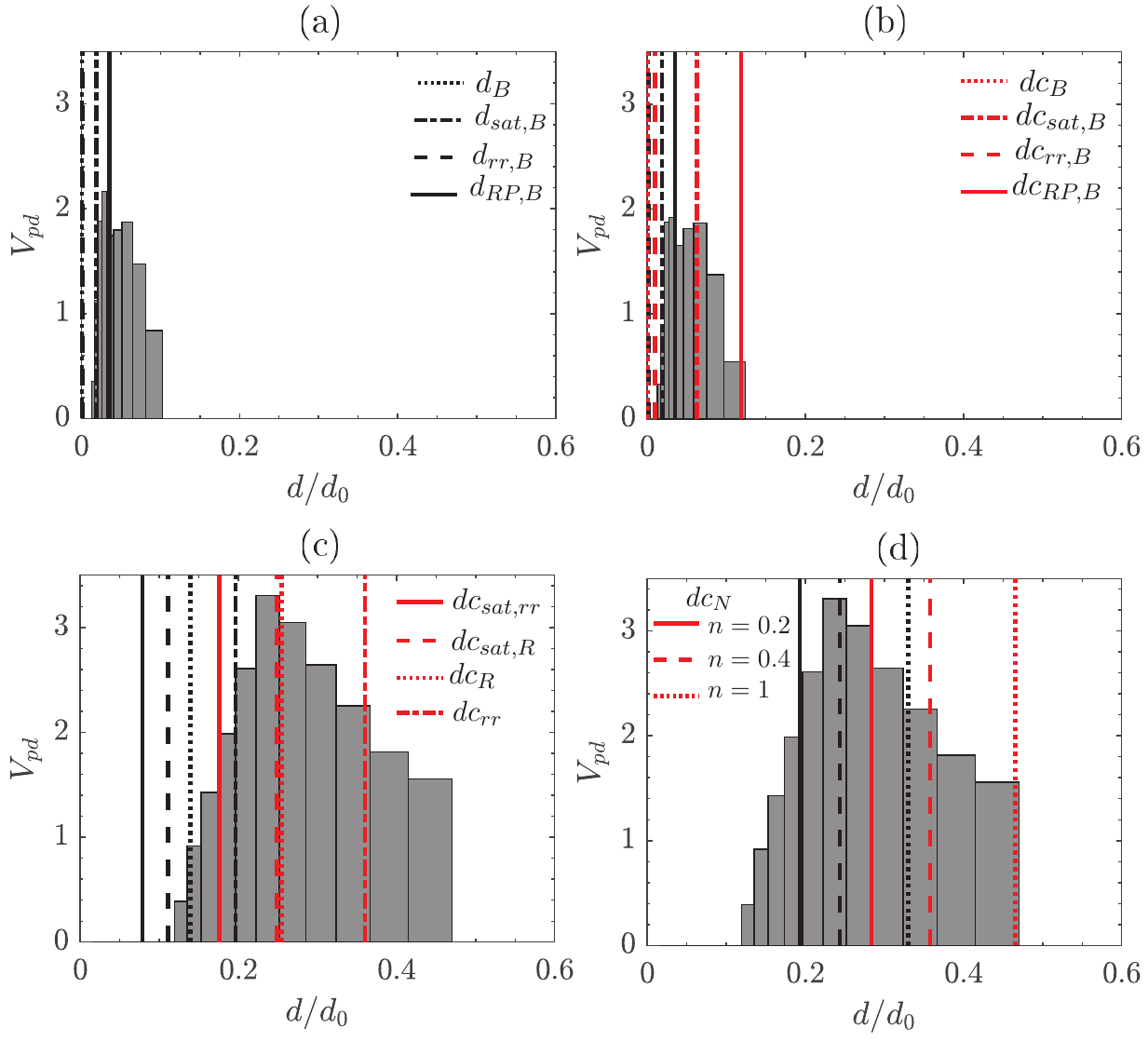} 
\caption{Comparison of experimental data with the theoretically predicted characteristic breakup sizes at (a) $\tau=0.30$, (b) $\tau=0.61$ and (c,d) $\tau=1.06$ for $\We = 34.8$ (dual-bag breakup). The panels (c) and (d) correspond to the rim and node fragmentations, respectively.}
\label{Appendix_fig2}
\end{figure}

The mode decomposition for the dual-bag breakup at $\We=34.8$ is presented in figure \ref{fig7} and the corresponding characteristic sizes for each mode are presented in figure \ref{Appendix_fig2}. It can be seen in figure \ref{fig7}(b) that at $\tau=0.3$, the parent bag ruptures, and the corresponding size distribution is presented in figure \ref{fig7}(e). We use Eq. (\ref{j:eq4}) to calculate the size distribution by only considering the child droplets from the parent bag breakup. At $\tau=0.61$, the core bag also breaks (figure \ref{fig7}c), and the combined contribution of the parent and core bags is shown in figure \ref{fig7}(f). For bag breakup, the analytical model predicts the experimentally observed size distributions and exhibits a mono-modal behavior. At $\tau=1.06$, we subtract the child droplets generated due to both bags from the total number of fragments to isolate the contributions of the rim and nodes (figure \ref{fig7}g). Both experimental results and analytical models indicate that child droplets resulting from rims and nodes exhibit mono-modal distributions. The combined contributions of the bags, rim, and nodes of the parent and core result in a bi-modal distribution, as shown in figure \ref{fig5}(b). The distinct behaviour observed in dual-bag breakup as compared to the single-bag case can be attributed to two factors. Firstly, in dual-bag fragmentation, the sizes of the child droplets produced from the core rim and parent nodes overlap with each other. Secondly, the separation between the characteristic sizes of the droplets from the parent rim and nodes is also significantly lower than that observed in the single-bag breakup. Our results also reveal that the total volume percentages of child droplets due to the parent bag and core bag are 18.9\% and 9\%, respectively. The combined contributions from the rim and nodes of the parent and core provide a total volume percentage of 72.12\%.

\subsection{Multi-bag breakup} \label{highWe}

\begin{figure}
\centering
\includegraphics[width=0.85\textwidth]{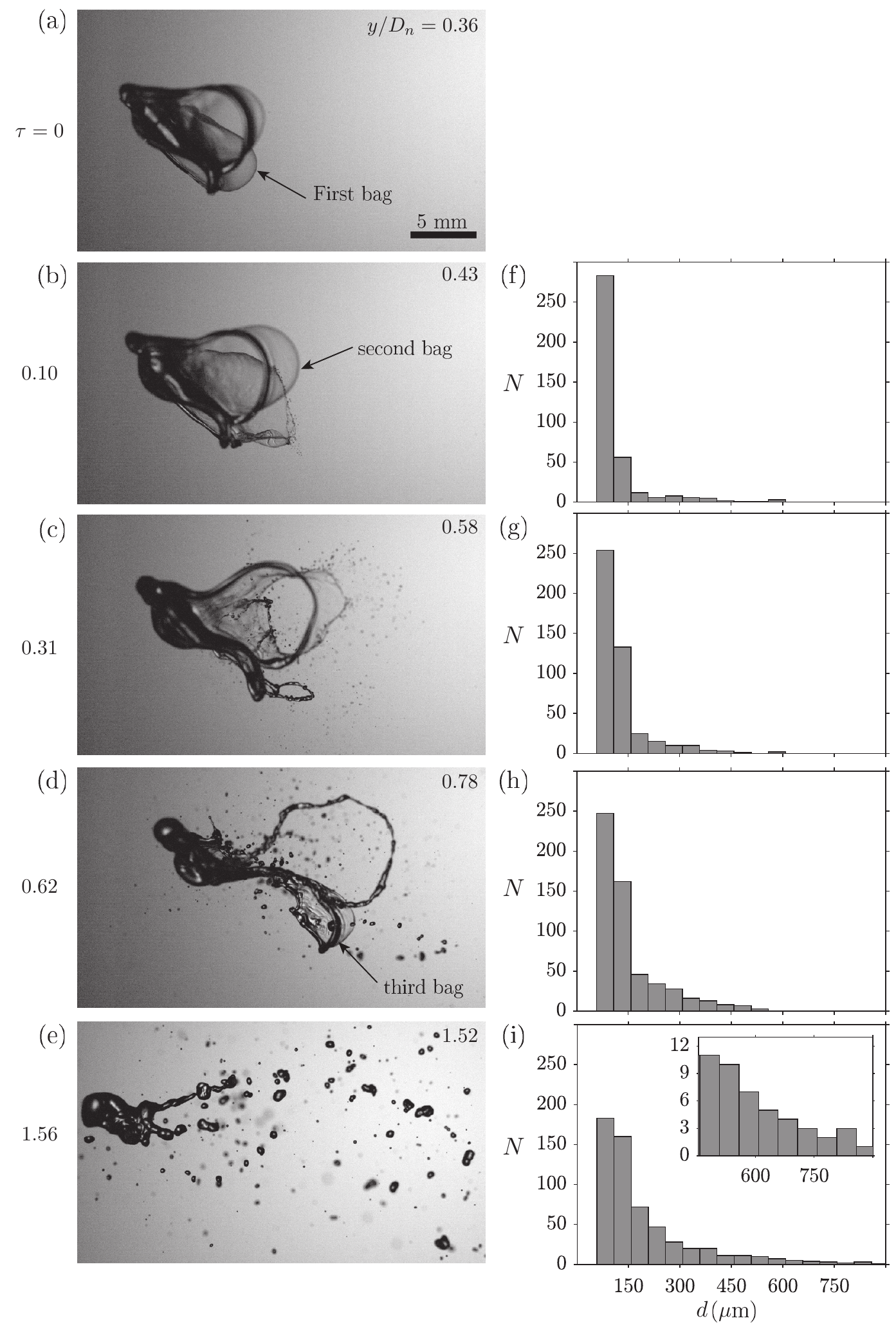}
\caption{Temporal evolution of a water droplet ($d_{0}=4.3$ mm) undergoing multi-bag fragmentation process ($\We = 40.15$).  Panels (a-e) depict shadowgraphy images at five instants corresponding to the maximum inflation of parent bag $(\tau=0)$, rupture of parent bag $(\tau=0.10)$, fragmenting of second bag $(\tau=0.31)$,  inflation of third bag $(\tau=0.62)$, and fragmentation of rim and nodes $(\tau=0.56)$, respectively. The dimensionless time $(\tau)$ and droplet position in the flow direction $(y/D_{n})$ are mentioned in each panel. The corresponding size distributions of the child droplets at $\tau=0.10$, 0.31, 0.62, and 1.06 are shown in panels (f-i), respectively.}
\label{fig8}
\end{figure}

In order to validate the analytical model at a high Weber number $(\We=40.15)$, we perform experiments for a bigger size of drop ($d_{0}=4.3$ mm). At this Weber number, the size of the undeformed core of the drop increases which results in multi-bag formation along the direction of the airstream. Figure \ref{fig8} depicts the temporal evolution of the droplet undergoing multi-bag fragmentation (panels a-e) and the corresponding size distributions (panels f-i) at $\We=40.15$. It can be seen that at $\tau=0$, two elongated bags are formed which are separated by a rib-like structure. The first bag ruptures at $\tau=0.1$ and produces child droplets in the range $0 \le d/d_0 < 450 ~ \mu$m. Subsequently, the second bag fragments and produces more tiny droplets at $\tau=0.31$. As the remaining undeformed core is exposed to the strong aerodynamic field, at $\tau=0.62$, the core further deforms and inflates into a bag-like structure (third bag), which also ruptures. Eventually, the rim and nodes associated with the first, second, and third bags fragment at $\tau=1.56$, resulting in larger child droplets of size, $d/d_0 >450 ~ \mu$m.

\begin{figure}
\centering
\includegraphics[width=0.45\textwidth]{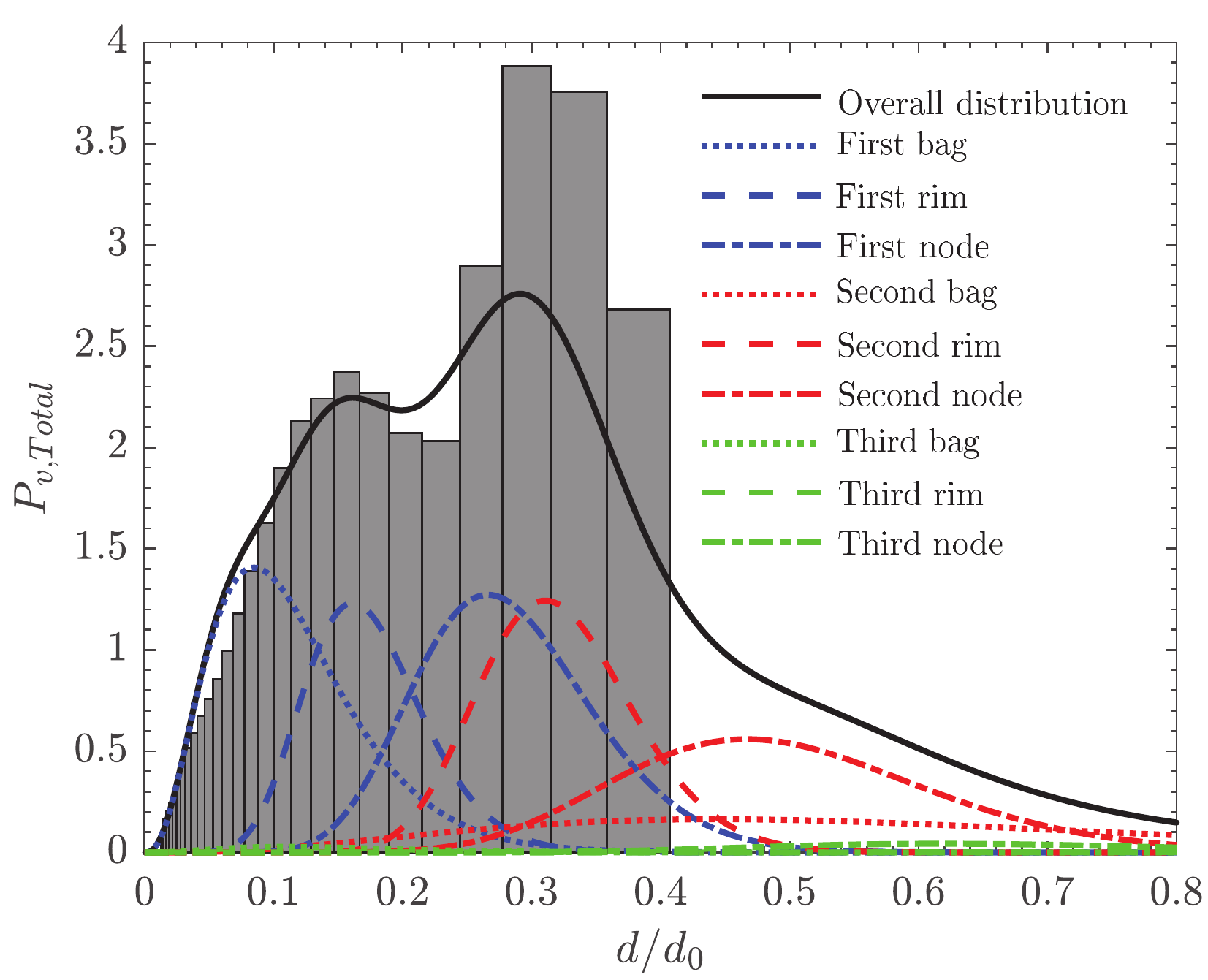}
\caption{Comparison of the multi-modal size distribution obtained from the experiments with the analytical predictions for $\We = 40.15$ at $\tau=1.56$.}
\label{fig9}
\end{figure}

\begin{figure}
\centering
\hspace{0.8cm}(a) \hspace{5.5cm}(b) \\
\includegraphics[height=0.3\textwidth]{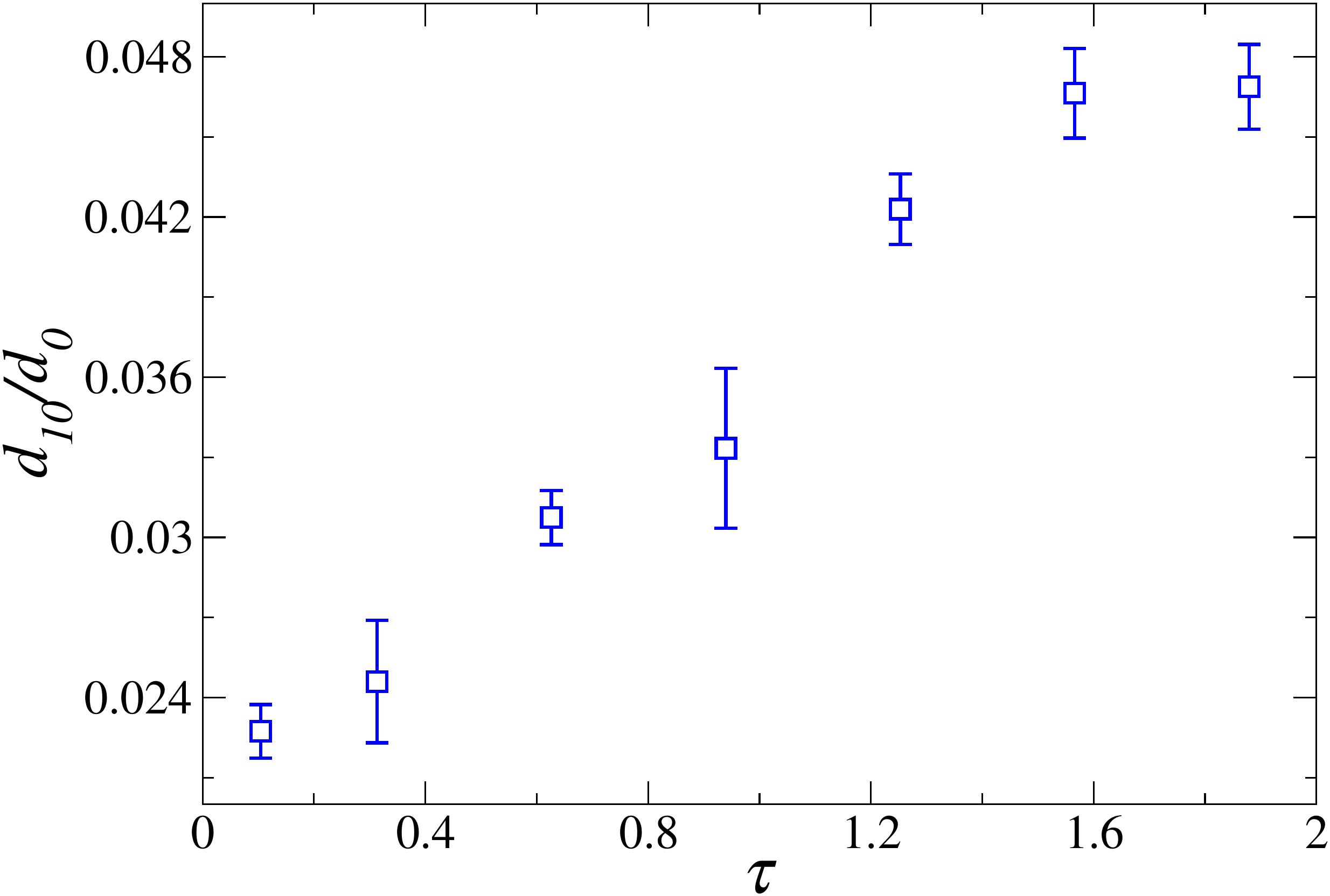} \ \includegraphics[height=0.3\textwidth]{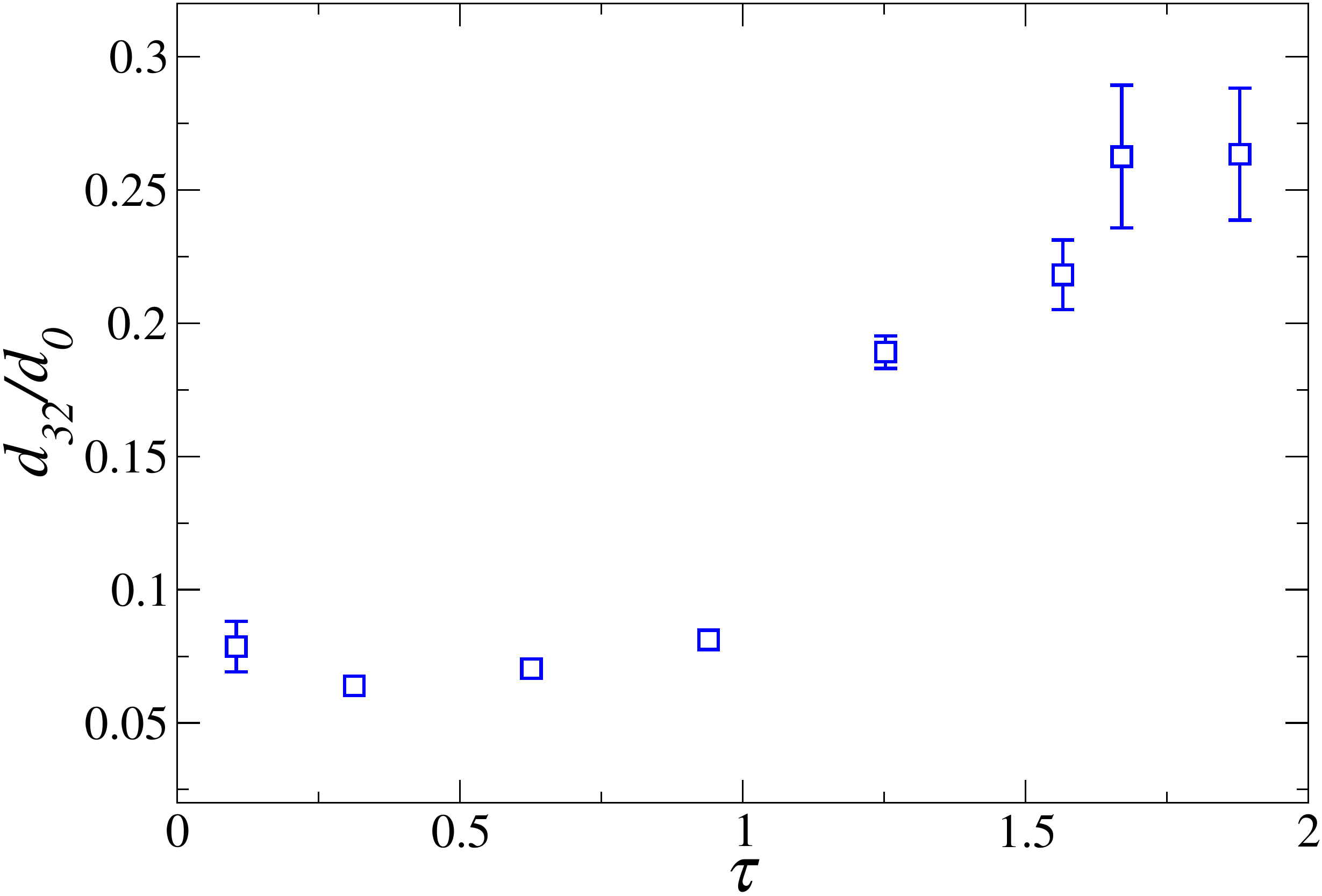} \\
\caption{Temporal variation of the normalised (a) number mean diameter $(d_{10}/d_0)$ and (b) Sauter mean diameter $(d_{32}/d_0)$ for $\We=40.15$. The error-bar represents the standard deviation obtained using three repetitions.}
\label{Appendix_fig7}
\end{figure}

The analytical model for the dual-bag fragmentation (discussed in \S\ref{sec31}) can be extended to the multi-bag fragmentation \citep{jackiw2022prediction}. The comparison of the size distribution obtained from the analytical model and experiments is shown in figure \ref{fig9} for the multi-bag breakup. Inspection of figure \ref{fig9} reveals that the first peak of the overall size distribution (at $d/d_{0} \approx 0.15$) is largely associated with the fragmentation of all the bags and the first rim. The second peak (at $d/d_{0} \approx 0.30$) results from the rims and nodes of the second and third bags. Among these bags, while the first bag contributes significantly to overall size distribution (18.5\%), the contributions from the second and third bags are 10.7\% and 0.5\%, respectively. The rest of the contributions are associated with the fragmentation of rims and nodes. The intricate breakup phenomenon at this high Weber number involves nine separate modes, yet yields a bi-model size distribution. The temporal variations of the normalised number mean diameter $(d_{10}/d_0)$ and Sauter mean diameter $(d_{32}/d_0)$ for $\We=40.15$ are plotted in figure \ref{Appendix_fig7}(a) and (b), respectively. We found that the trend is similar to that observed for $\We=12.6$ and 34.8. However, the time scale of the fragmentation is larger in the case of the multi-bag breakup.

\begin{figure}
\centering
\includegraphics[width=0.45\textwidth]{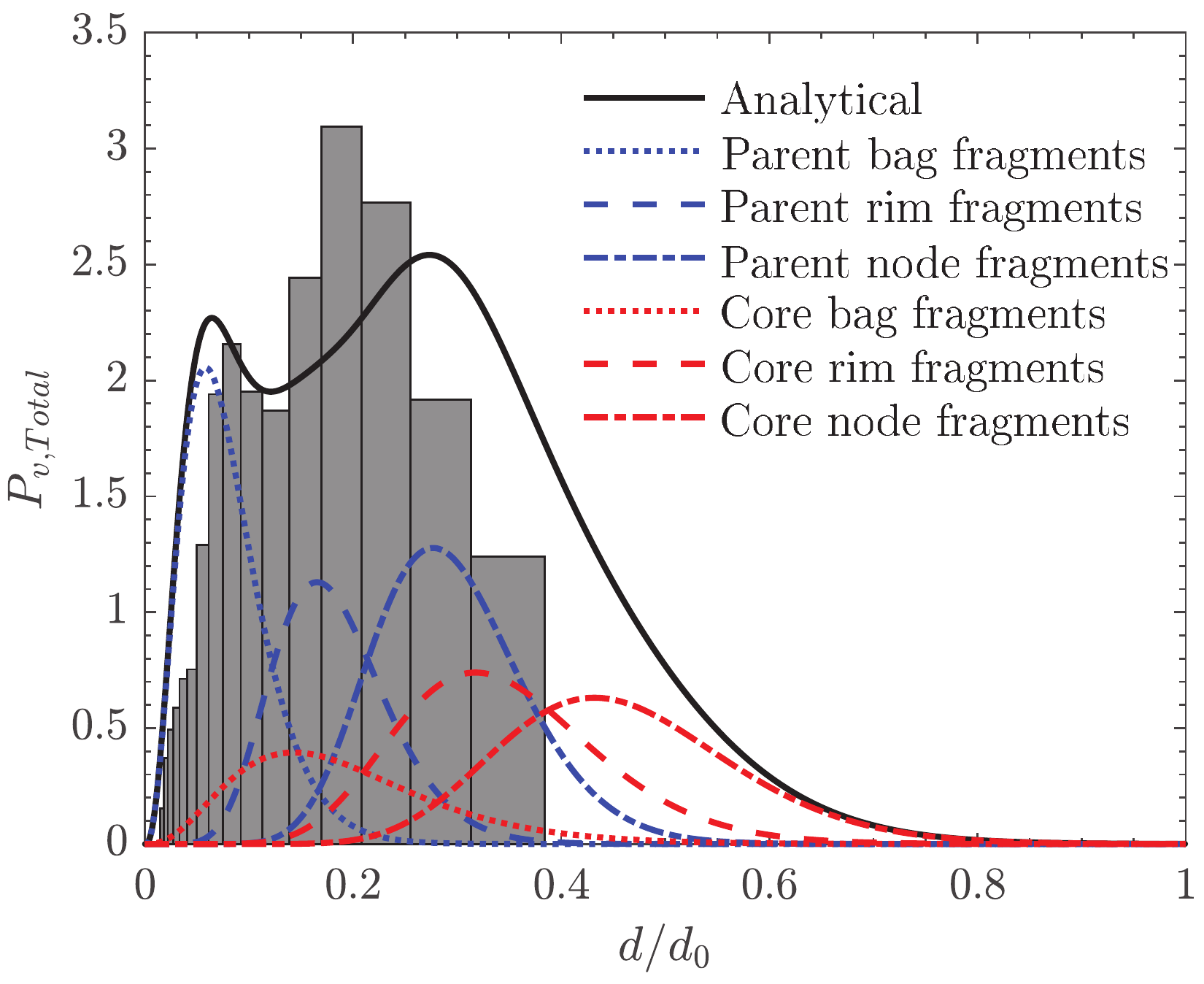}
\caption{Comparison of the analytical model with the experimental data for a water drop of initial diameter $d_{0}= 4.3$ mm and $\We = 30.0$. This exhibits a dual-bag breakup.}
\label{Appendix_fig4}
\end{figure}

\begin{figure}
\centering
\hspace{0.8cm}(a) \hspace{5.5cm}(b) \\
\includegraphics[width=0.462\textwidth]{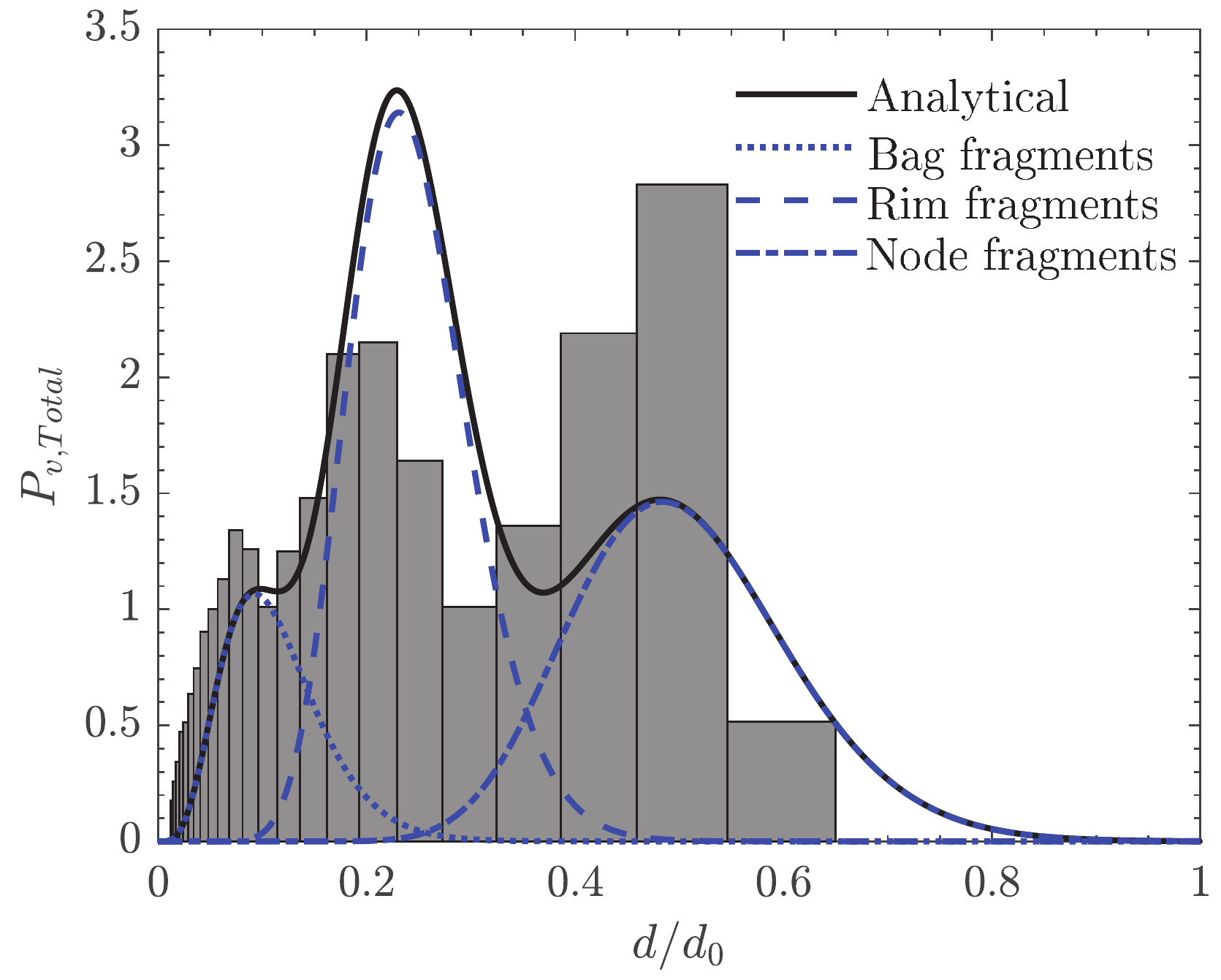} \includegraphics[width=0.45\textwidth]{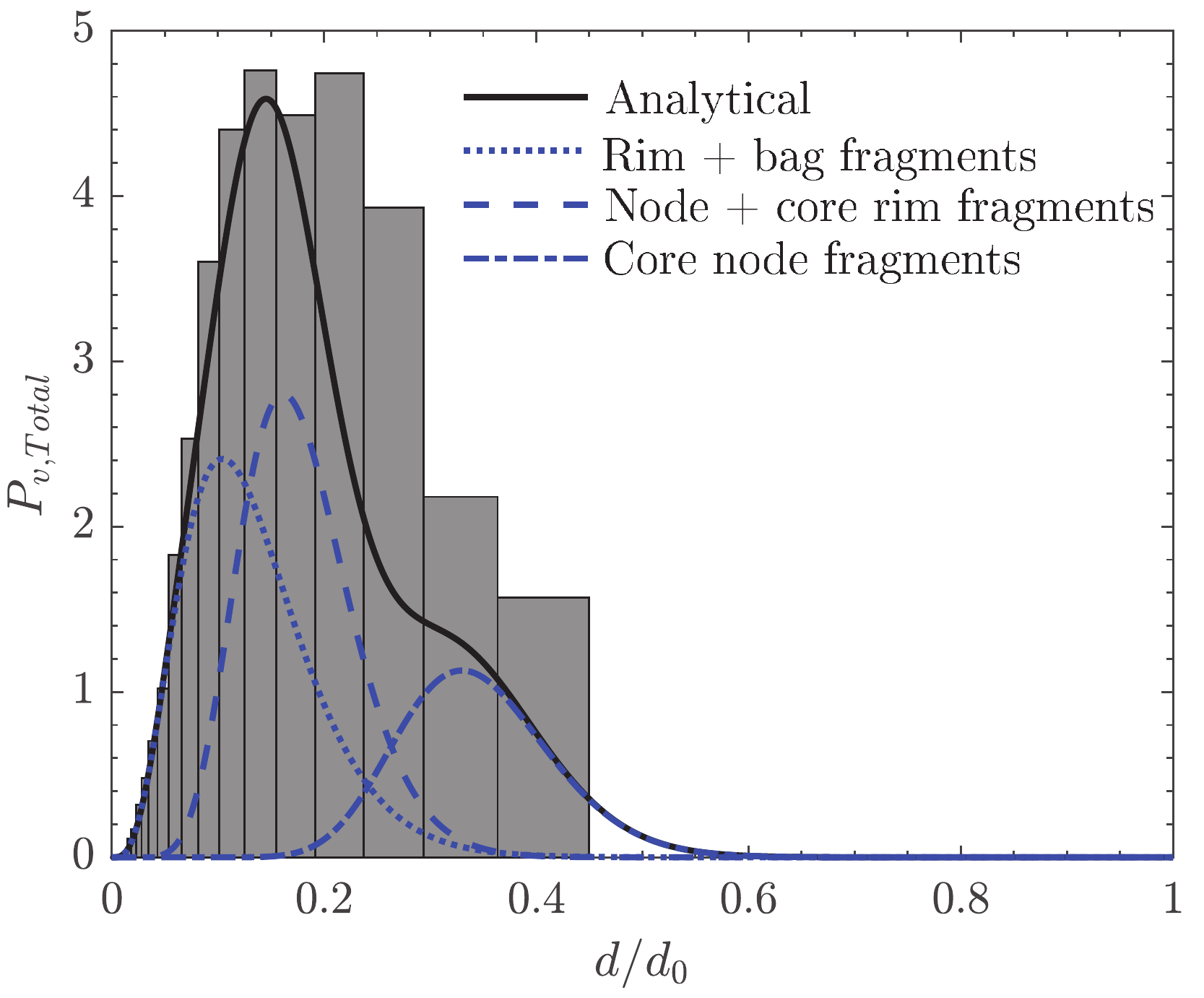}
\caption{Comparison of the analytical model with the experimental data taken from \cite{guildenbecher2017characterization} for an ethanol drop of initial diameter $d_{0}=2.7$ mm at (a) $\We=13.8$ (single-bag breakup) and (b) $\We=55.3$ (sheet-thinning breakup).}
\label{Appendix_fig5}
\end{figure}

Additionally, in order to check the validity of the model, we investigate the droplet size distributions for three more Weber numbers, namely, $\We=30.0$ (water droplet of initial diameter $d_{0}=4.3$ mm), $\We=13.8$ (ethanol drop of initial diameter $d_{0}=2.7$ mm) and $\We=55.3$ (ethanol drop of initial diameter $d_{0}=2.7$ mm). These droplets undergo bag, dual-bag and sheet-thinning breakups at $\We=13.8$, 30.0 and 55.3, respectively. While we have performed the experiments for $\We=30.0$, the data for $\We=13.8$ and 55.3 have been taken from \citet{guildenbecher2017characterization}. The comparisons of the analytical model with the experimental data for different values of $\We$ are presented in figure \ref{Appendix_fig4} and figure \ref{Appendix_fig5}(a,b). It can be observed that the analytical model convincingly predicts the size distribution of the child droplets for a range of Weber numbers.

\section{Concluding remarks}
\label{sec:conc}
We investigate the droplet size distribution owing to the dual-bag fragmentation of a water drop and compare it to the single-bag breakup using shadowgraphy and deep learning-based digital in-line holography. We find that the Sauter mean diameter of the child droplets in dual-bag fragmentation is smaller than that observed in the single-bag breakup. Secondly, the size of the child droplets resulting from the fragmentation of the parent drop is smaller than the core drop. Additionally, we utilize the analytical model developed by \cite{jackiw2022prediction} to predict the volume probability density of the child droplets resulting from the dual-bag fragmentation and compare it with our experimental result. Interestingly, despite having six distinct breakup phenomena, the dual-bag breakup exhibits a bi-modal distribution in contrast to the single-bag breakup, which undergoes a tri-modal distribution. We have estimated the temporal evolution of the child droplet production to quantitatively show the decomposition of the model into the initial and core breakups. We further show that the analytical model predicts the droplet size distribution for a range of Weber numbers. \\

\vspace{2mm}
\noindent{\bf Declaration of Interests:} The authors report no conflict of interest. \\

\noindent{\bf Acknowledgement:} {L.D.C. and K.C.S. thank the Science \& Engineering Research Board, India for their financial support through grants SRG/2021/001048 and CRG/2020/000507, respectively. We also thank IIT Hyderabad for the financial support through grant IITH/CHE/F011/SOCH1 and start-up grant (for L.D.C.). S.S.A. also thanks the PMRF Fellowship.}


\end{document}